\documentclass[fleqn,usenatbib]{mnras}
\usepackage{import}

\import{./frontmatter/}{preamble.sty}
\def \aj {{\rm {AJ}}}
\def \araa {{\rm {ARA\&A}}}
\def \apj {{\rm {ApJ}}}

\def \apjs {{\rm {ApJS}}}
\def \apjl {{\rm {ApJL}}}

\def \aap {{\rm {A\&A}}}

\def \mnras {{\rm {MNRAS}}}

\def \pasp {{\rm {PASP}}}

\def \nat {{\rm {Nat}}}
\def \kms {{\rm km\,s$^{-1}$}}

\def \sol {{\rm M$_\odot$}}

\def \arcsec {{\rm $^{\prime\prime}$}}
\def \micron{\hbox{$\upmu$m}}

\def \cmcb {{\rm cm$^{-3}$}}

\def \ntwoh {{\rm N$_2$H$^+$}}

\def \tcooz {{\rm $^{13}$CO\,($1-0$)}}
\def \ntwohoz {{\rm N$_2$H$^+$\,($1-0$)}}

\newcommand{\note}[1]{\textcolor{black}{ #1}}

\defcitealias{rathborne_2006}{R06}
\defcitealias{henshaw_2016c}{H16}
\defcitealias{henshaw_2016d}{H17}
\defcitealias{liu_2018}{L18}

\defcitealias{larson_1981}{L81}
\defcitealias{kauffmann_2010}{KP10}
\defcitealias{krumholz_2008}{KM08}
\defcitealias{caselli_1995b}{CM15}
\defcitealias{kauffmann_2013}{K13}
\defcitealias{peretto_2013}{P13}
\defcitealias{sanchezmonge13}{SM13}

\title[ALMA-IRDC: dense gas mass distribution]{ALMA-IRDC: the dense gas mass distribution from (infrared dark) molecular cloud (>1pc) to massive star-forming cores scales (0.01pc)}

\title[ALMA-IRDC: dense gas mass distribution]{ALMA-IRDC: Dense gas mass distribution from cloud to core scales}

\author[A.~T.~Barnes et al.]{A.~T.~Barnes,$^{1}$\thanks{E-mail: ashleybarnes.astro@gmail.com}
   J.~D.~Henshaw,$^{2}$
   F.~Fontani,$^{3,4}$
   J.~E.~Pineda,$^{3}$
   G.~Cosentino,$^{5}$
   J.~C.~Tan,$^{5,6}$ \and
   P.~Caselli,$^{3}$ 
   I.~Jiménez-Serra,$^{7}$ 
   C.~Y.~Law,$^{5}$ 
   A.~Avison,$^{8,9}$
   F.~Bigiel,$^{1}$
   S.~Feng,$^{10,11,12}$ \and
   S.~Kong,$^{13}$ 
   S.~N.~Longmore,$^{14}$
   L.~Moser,$^{1}$
   R.~J.~Parker,$^{15}$\thanks{Royal Society Dorothy Hodgkin Fellow.}
   Á.~Sánchez-Monge,$^{16}$ \and 
   and K.~Wang$^{17}$
   \\
    $^{1}$ Argelander-Institut f\"{u}r Astronomie, Universit\"{a}t Bonn, Auf dem H\"{u}gel 71, 53121, Bonn, Germany \\
    $^{2}$ Max Planck Institute for Astronomy, K\"{o}nigstuhl 17, D-69117 Heidelberg, Germany\\
    $^{3}$ Max-Planck-Institut f\"{u}r extraterrestrische Physik, Giessenbachstrasse 1, 85748 Garching bei M\"{u}nchen, Germany \\
    $^{4}$ INAF Osservatorio Astrofisico di Arcetri, Largo E. Fermi 5, 50125 Florence, Italy \\
    $^{5}$ Dept. of Space, Earth and Environment, Chalmers University of Technology, SE-412 96 Gothenburg, Sweden \\
    $^{6}$ Department of Astronomy, University of Virginia, 530 McCormick Road Charlottesville, 22904-4325 USA \\
    $^{7}$ Centro de Astrobiolog\'{i}a (CSIC/INTA), Instituto Nacional de T\'{e}cnica Aeroespacial, 28850 Torrej\'{o}n de Ardoz, Madrid, Spain \\
    $^{8}$ UK ALMA Regional Centre Node, Manchester, M13 9PL, UK\\ 
    $^{9}$ Jodrell Bank Centre for Astrophysics, Alan Turing Building, School of Physics and Astronomy, The University of Manchester, Manchester, M13 9PL, UK \\ 
    $^{10}$ National Astronomical Observatories, Chinese Academy of Science, Beijing 100101, People's Republic of China \\ 
    $^{11}$ Academia Sinica Institute of Astronomy and Astrophysics, No.1, Section 4, Roosevelt Road, Taipei 10617, Taiwan, Republic of China \\ 
    $^{12}$ National Astronomical Observatory of Japan, National Institutes of Natural Sciences, 2-21-1 Osawa, Mitaka, Tokyo 181-8588, Japan \\ 
    $^{13}$ Steward Observatory, University of Arizona, Tucson, AZ 85719, USA \\ 
    $^{14}$ Astrophysics Research Institute, Liverpool John Moores University, 146 Brownlow Hill, Liverpool L3 5RF, UK \\ 
    $^{15}$ Department of Physics and Astronomy, The University of Sheffield, Hicks Building, Hounsfield Road, Sheffield, S3 7RH, UK \\ 
    $^{16}$ I. Physikalisches Institut, Universität zu Köln, Zülpicher Str. 77, 50937 Köln, Germany \\ 
    $^{17}$ Kavli Institute for Astronomy and Astrophysics, Peking University, 5 Yiheyuan Road, Haidian District, Beijing 100871, China 
    }

\date{Accepted 2021 March 6. Received 2021 February 12; in original form 2020 October 9.}

\pubyear{2020}

\begin{document}

\label{firstpage}
\pagerange{\pageref{firstpage}--\pageref{lastpage}}
\maketitle

\begin{abstract}
Infrared dark clouds (IRDCs) are potential hosts of the elusive early phases of high-mass star formation (HMSF). Here we conduct an in-depth analysis of the fragmentation properties of a sample of 10 IRDCs, which have been highlighted as some of the best candidates to \note{study} HMSF within the Milky Way. To do so, we have obtained a set of large mosaics covering these IRDCs with ALMA at band 3 (or 3\,mm). These observations have a high angular resolution ($\sim$\,3\arcsec; $\sim$\,0.05\,pc), and high continuum and spectral line sensitivity ($\sim$\,0.15\,mJy\,beam$^{-1}$ and $\sim$\,0.2\,K per 0.1\,\kms\ channel at the \ntwohoz\ transition). From the dust continuum emission, we identify 96 cores ranging from low- to high-mass ($M=3.4 - 50.9$\sol) that are gravitationally bound ($\alpha_\mathrm{vir} = 0.3 - 1.3$) and which would require magnetic field strengths of $B=0.3 - 1.0$\,mG to be in virial equilibrium. We combine these results with a homogenised catalogue of literature cores to recover the hierarchical structure within these clouds over four orders of magnitude in spatial scale (0.01\,pc -- 10\,pc). Using supplementary observations at an even higher angular resolution, we find that the smallest fragments (<\,0.02\,pc) within this hierarchy do not currently have the mass and/or the density required to form high-mass stars. Nonetheless, the new ALMA observations presented in this paper have facilitated the identification of 19 (6 quiescent and 13 star-forming) cores that retain >16\,\sol\ without further fragmentation. These high-mass cores contain trans-sonic non-thermal motions, are kinematically sub-virial, and require moderate magnetic field strengths for support against collapse. The identification of these potential sites of high-mass star formation represents a key step in allowing us to test the predictions from high-mass star and cluster formation theories.
\end{abstract}





\section{Introduction}\label{sec:intro}

\begin{figure*}
\centering
    	\includegraphics[width=1\textwidth]{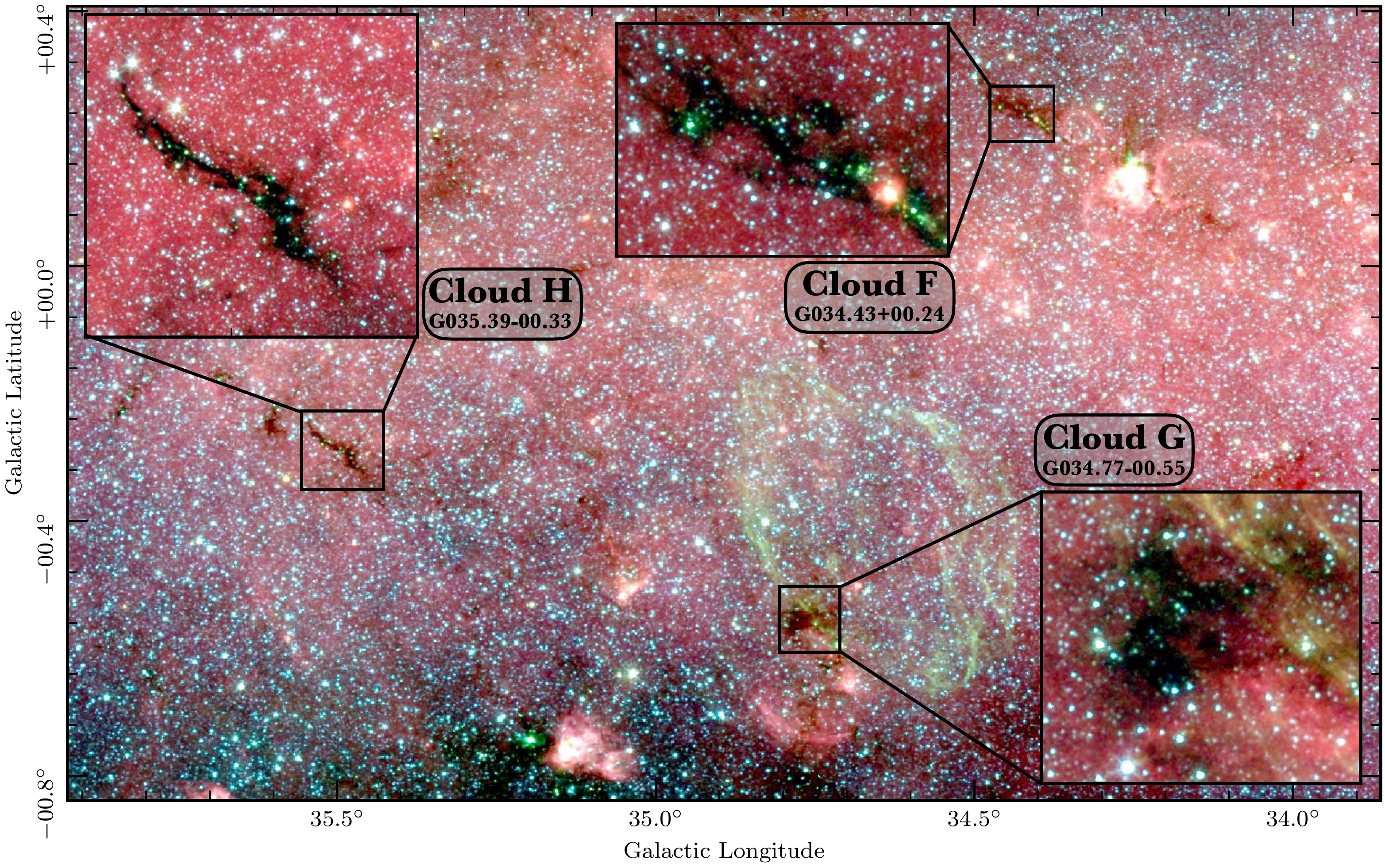}
    	\caption{A three colour image of the Galactic plane where several infrared dark clouds (IRDCs) can be seen as dark extinction features. In this image, red is 8\,\micron, green is 5.8\,\micron\ and blue is 4.5\,\micron\ emission from the {\it Spitzer} GLIMPSE survey \citep{carey_2009}. Labelled are three IRDCs that are investigated within this work: Clouds F, G and H (or G034.43+00.24, G034.77-00.55, G035.39-00.33). The panels show zoom-ins of these IRDCs for more detail.}
    	\label{fig:rgb_map}
\end{figure*}

High-mass ($>8$\sol) stars are of great astrophysical importance due to the large amounts of energy and momentum, along with the production of heavy elements, that they inject into the interstellar medium throughout their short lifetimes \citep{zinnecker07}. The later evolutionary stages, once the massive star has formed, have been well studied, and their role in driving the evolution of their host environment, and even the host galaxy, is relatively well understood. However, despite ongoing efforts, the earliest evolutionary stages, during the formation process of these massive stars, are not nearly as well constrained (e.g. \citealp{tan_2014, motte18}). Studies of the initial conditions of high-mass star-forming regions are required to unveil their formation mechanisms, before the disruptive effects of protostellar feedback disperse molecular clouds on a short timescale (e.g. \citealp{kruijssen19a, barnes20b, chevance20c,chevance20}). This, however, first necessitates the identification of molecular clouds with sufficient mass and density, which currently exhibit a low star formation activity.

Infrared dark clouds (IRDCs) are a group of molecular clouds, the massive of which present promising candidates to study these initial conditions of high-mass star formation. These were initially identified with the {\it Infrared Space Observatory} ({\it ISO}; 15\,\micron; \citealp{perault_1996}) and the {\it Midcourse Space Experiment} ({\it MSX}; 7 to 25\,\micron; \citealp{egan_1998}) as regions of strong mid-infrared extinction against the background Galactic emission, highlighting that they must contain substantial dust column densities. Subsequent work \note{found that} IRDCs \note{can be} cold ($<$\,20\,K; \citealp{pillai_2006, ragan_2011}), high-mass ($\sim$\,10$^{3-5}$\,\sol; \citealp{rathborne_2006, longmore_2012, kainulainen_2013}), have large column densities ({\it N}(H$_2$)\,$\sim$\,$10^{22-25}$\,cm$^{-2}$; \citealp{egan_1998, carey_1998, simon_2006a, vasyunina_2009}), and have high mean number densities ({\it n}(H$_2$)\,$\sim$\,10$^{3-5}$cm$^{-3}$; e.g. \citealp{peretto_2010a, peretto_2010b, hernandez_2011, butler_2012}). Of particular importance, IRDCs \note{can contain} large reservoirs of relatively pristine gas, which has not been influenced by star formation, as inferred from their chemical composition (e.g. \citealp{miettinen_2011, gerner_2015, barnes_2016, kong_2016}).

This is the first in a series of papers, which aims to conduct an in-depth assessment of the initial physical, chemical and kinematic conditions for massive star/cluster formation across a sample of IRDCs using a suite of recently obtained Atacama Large Millimeter/submillimeter Array (ALMA) observations. The 10 cloud sample has been singled out by the stringent selection process summarised below, as being \note{particularly good} candidates in which to study the initial conditions of massive star formation; see Figure\,1 from \citealp{tan_2014} for comparison of their properties to wider molecular cloud population. Firstly, the cloud sample was initially identified along with $\sim$11,000 other candidate IRDCs in the study of \citet{simon_2006a}, which showed extended structure silhouetted against diffuse background emission. \citet{simon_2006b} then investigated the global properties of a sub-sample of these $\sim$11,000 clouds that were extended, had high extinctions, and were covered by the Galactic Ring Survey (a survey of \tcooz\ molecular line emission; \citealp{jackson_2006}). \citet{rathborne_2006} then investigated the clump properties within 38 of these clouds, selecting those which had known kinematic distance estimates \citep{simon_2006b}. Finally, \citet{butler_2009, butler_2012} and \citet{kainulainen_2013} studied the near- and mid-infrared extinction properties within 10 clouds of the \citet{rathborne_2006} sample, which were specifically chosen as being relatively nearby and massive. The properties of this cloud sample is given in Table\,\ref{tab:cloudprops}. Figure\,\ref{fig:rgb_map} shows a mid-infrared image covering $\sim$2 degrees of the Galactic plane, where the positions of three clouds from our sample can be clearly seen as dark extinction features. 

In this first Paper, we investigate how the large, dense and pristine mass reservoirs available within the cloud sample fragment down to the scales of individual (massive) star-forming cores ($\sim$\,0.01\,pc or $\sim$\,1000\,au). This study is motivated by the need for observational constraints on the hierarchical mass distribution of IRDCs for testing the different theories of massive star formation. In many studies over the last decade, this was boiled down to differentiating between the predictions of core-accretion models (e.g. \citealp{mckee_2003}), where massive stars are born from the collapse of a massive core where small scale fragmentation is suppressed, and competitive accretion models (e.g. \citealp{bonnell01,bonnell_2004}), where the gas is highly fragmented into many thermal Jeans mass cores that form low-mass protostars, which then competitively accrete from the host clump environment. Although, other theories have emerged more recently (e.g. \citealp{vazquezsemadeni19, padoan20}). This paper provides the first in a suite of observational tests that aim at differentiating between these various prescriptions for high-mass star formation. 

This work is organised as the following. In section\,\ref{sec:observations} we give details of the ALMA observations of the 3\,mm dust continuum and the \ntwohoz\ line transition, which is thought to trace cold and dense molecular gas (e.g. \citealp{pety_2017,kauffmann_2017,barnes20a}). The results of the core identification and the calculation of their physical and dynamical properties are presented in section\,\ref{sec:results}. In section\,\ref{sec:analysis}, we outline the procedure used to create the homogenised literature core catalogue, where the same set of physical assumptions is used to recalculate previously published radio continuum core catalogues covering our cloud sample. This homogenised core catalogue is then analysed and compared to various scaling relations. Moreover, in section\,\ref{sec:analysis}, we link cores identified here to the cores from the literature catalogue, which allows us to identify several potential sites of high-mass star formation, and follow the physical properties of the determined hierarchical structure from the cloud ($\sim1$\,pc) down to individual star-forming core scales ($\sim0.01$\,pc). This work is then summarised in section\,\ref{sec:conclusions}. The appendix gives an example of the core and homogenised core catalogues, which can be found in full, machine-readable format online.  
 
\begin{table*}
\centering
\caption{Table of the global properties across the cloud sample. Shown in columns are the cloud names from \citet{butler_2012}, IDs from \citet{rathborne_2006}, the systemic velocity ($v_\mathrm{sys}$), cloud kinematic distances ($D_\mathrm{cl}$), effective radii ($R_\mathrm{eff,cl}$), masses determined from near- and mid- infrared extinction mapping ($M_\mathrm{cl}$), velocity dispersions from $^{13}$CO\,(1-0) emission ($\sigma_\mathrm{cl}$), virial parameters ($\alpha_\mathrm{vir,cl}$), have all been taken from \citet[][Table 1]{kainulainen_2013}. Also given is the mean {\it Herschel} derived dust temperature measured over the ALMA coverage ($T_\mathrm{dust,cl}$), the non-thermal velocity dispersion and sonic Mach number ($\sigma_\mathrm{NT,cl}$ and $\mathcal{M}_\mathrm{NT,cl}$; section\,\ref{sec:dynprops}), and the number of cores identified in the ALMA continuum observations (section\,\ref{sec:coreident}).}

    \begin{tabular}{cccccccccccc}
    \hline \hline
    Cloud & ID & $\mathrm{v}_\mathrm{sys}$ & $D_\mathrm{cl}$ & $R_\mathrm{eff,cl}$ & $M_\mathrm{cl}$ & $\sigma_\mathrm{cl}$ & $\alpha_\mathrm{vir,cl}$ & $T_\mathrm{dust,cl}$ & $\sigma_\mathrm{NT,cl}$ & $\mathcal{M}_\mathrm{NT,cl}$ & $n_\mathrm{c}$ \\
    & ``G(longitude)(latitude)'' & \kms & $\mathrm{pc}$ & $\mathrm{pc}$ & $\mathrm{M_{\odot}}$ & $\mathrm{km\,s^{-1}}$ & - & $\mathrm{K}$ & $\mathrm{km\,s^{-1}}$ & - & $\#$ \\
     \hline
  
    Cloud A & G018.82-00.28 & 59-69 & 4800 & 10.4 & 18500.0 & 2.04 & 1.4 & 18.3 & 2.0 & 7.9 & 8 \\
    Cloud B & G019.27+00.07 & 22-32 & 2400 & 2.71 & 2200.0 & 1.6 & 2.2 & 17.8 & 1.6 & 6.3 & 10 \\
    Cloud C & G028.37+00.07 & 73-83 & 5000 & 15.4 & 53200.0 & 3.72 & 2.4 & 17.7 & 3.7 & 14.8 & 16 \\
    Cloud D & G028.53-00.25 & 81-91 & 5700 & 16.9 & 74300.0 & 1.85 & 0.5 & 16.9 & 1.8 & 7.5 & 16 \\
    Cloud E & G028.67+00.13 & 75-85 & 5100 & 11.5 & 28700.0 & 4.32 & 1.1 & 19.1 & 4.3 & 16.5 & 4 \\
    Cloud F & G034.43+00.24 & 52-62 & 3700.0 & 3.5 & 4460.0 & 3.62 & 1.3 & 20.0 & 3.6 & 13.5 & 20 \\
    Cloud G & G034.77-00.55 & 35-45 & 2900 & 3.06 & 3300.0 & 3.28 & 4.7 & 19.8 & 3.3 & 12.3 & 0 \\
    Cloud H & G035.39-00.33 & 38-48 & 2900 & 9.69 & 16700.0 & 2.03 & 0.7 & 19.5 & 2.0 & 7.6 & 10 \\
    Cloud I & G038.95-00.47 & 38-48 & 2700 & 3.73 & 2700.0 & 1.65 & 1.2 & 18.0 & 1.6 & 6.4 & 9 \\
    Cloud J & G053.11+00.05 & 17-27  & 1800 & 0.755 & 200.0 & 0.96 & 1.5 & 19.0 & 0.9 & 3.5 & 3 \\


    \hline \hline
    \end{tabular}

\label{tab:cloudprops}
\end{table*}
\section{Observations}\label{sec:observations}

To investigate the dense gas properties within the IRDC sample, we have acquired high-angular resolution dust continuum and molecular line observations with ALMA as part of the projects: 2017.1.00687.S and 2018.1.00850.S (PI: A.T. Barnes). The observations made use of the Band 3 receiver, which was configured to obtain high spectral resolution observations (0.1\,\kms\ or 30.518\,kHz) of \ntwohoz\ centred at $\sim$93\,GHz, and a broad continuum bandwidth of $\sim$\,4\,GHz. Complementary observations were made in the C43-1 12\,m array configuration (baselines of 15 to 314\,m) and 7\,m (ACA) array (baselines of 8 to 48\,m). Single dish total power observations were also performed for the molecular lines. This observational setup was chosen to be directly comparable to the Plateau de Bure interferometer (NOEMA precursor) observations of the northern portion of one of the clouds in the sample (Cloud H; \citealp{henshaw_2014, henshaw_2016c}). 

The 12\,m and 7\,m array observations were reduced and imaged by the {\sc casa-pipeline} (version: 5.4.0-70). In this work, we wish to make a direct comparison between the continuum and molecular line emission, and, hence, only make use of the 12\,m and 7\,m observations that are available for both spectral configurations. The mosaic images from these arrays were combined with the {\sc feather} function in CASA (version 4.7.0; \citealp{mcmullin07}) with the default parameter set (i.e. effective dish size, single-dish scaling, and low-pass filtering of the single-dish observations). We present an in-depth analysis of the 12\,m and 7\,m combination using the {\sc feather} function in appendix\,\ref{apendix:cleanfeather} (see Figures\,\ref{fig:cleanmaps} and \ref{fig:cleanspec}), and deem this method to be accurate to within the underlying systematic uncertainties on the properties calculated within this work. The maximum recoverable scale within the combined images is set by the size of the smallest 7\,m baseline of 8\,m or 70\arcsec\ at 93.2GHz. The average angular beam size achieved within the combined observations for both the continuum and molecular line observations is 2.9\arcsec, which, across the sample with distances ranging $1.8 - 5.7$\,kpc, is equivalent to a projected length scale range of $0.05-0.1$\,pc. 

The average continuum (4\,GHz bandwidth) sensitivity achieved within the combined images with and without primary beam correction is 0.08\,mJy\,beam$^{-1}$ and 0.15\,mJy\,beam$^{-1}$, respectively. We quote both here as the continuum images without primary beam correction are used later in this work in the source identification routine, whilst the images with primary beam correction are then used to measure core fluxes and any calculated physical properties. The corresponding mass sensitivity of the primary beam corrected image is $\sim$\,1\,\sol\ (assuming a dust temperature of 20\,K; see equation\,\ref{equ:mass}). The average molecular line (30.518\,kHz/0.1\,\kms\ channel width) sensitivity is 15mJy\,beam$^{-1}$ or 0.2\,K, which was chosen to allow a significant detection of the isolated hyperfine component of \ntwoh\,(J,F${_1}$,F = 1,0,1 $\rightarrow$ 0,1,2) across the sample (93.1762522\,GHz; \citealp{caselli_1995a, pagani_2009}).\footnote{The isolated hyperfine component is used to get an accurate measurement of the line-width in the dynamical analysis of this work (section\,\ref{sec:dynprops}), as, unlike the main hyperfine component, this is unlikely to merge with other hyperfine components and suffers from lower fractional optical depth (e.g. see \citealp{henshaw_2014, barnes_2018}).} The beam size and sensitivity information of the final continuum map, and N$_2$H$^+$ cube is presented in Table\,\ref{tab:obsprops}.

Ultimately, these observations allow us to accurately recover the dense $0.1$\,pc scale core population across a sample of IRDCs, and investigate their spatial distributions, and physical, kinematic and chemical properties. An immediate follow-up work in the series of papers using the ALMA observations presented here, is the investigation of the $^{14}$N/$^{15}$N fraction observed in N$_2$H$^+$ \citep{fontani_2020inprep}. 


\begin{table}
\centering
    \caption{Table of the observational properties. Columned is the minor and major beam size, and mean rms value within the ALMA 3\,mm continuum map and \ntwohoz\ cube. The values of the mean continuum rms shown with and without parentheses have been determined using the maps with and without primary beam correction, respectively (see section\,\ref{sec:coreident}). The mean \ntwohoz\ cube rms values have been determined within a 0.1\,\kms\ channel.}
    \label{tab:obsprops}
    \begin{tabular}{ccccccccc}
    \hline \hline
    Cloud & \multicolumn{3}{c}{Continuum map} & \multicolumn{3}{c}{N$_2$H$^+$ cube} \\
     & $\theta_\mathrm{min}$ & $\theta_\mathrm{maj}$ & rms & $\theta_\mathrm{min}$ & $\theta_\mathrm{maj}$ & rms \\
    
     & $\mathrm{{}^{\prime\prime}}$ & $\mathrm{{}^{\prime\prime}}$ & $\mathrm{mJy\,beam^{-1}}$ & $\mathrm{{}^{\prime\prime}}$ & $\mathrm{{}^{\prime\prime}}$ & $\mathrm{K}$ \\
    \hline
    Cloud A & 2.54 & 3.15 & 0.08 (0.16) & 2.91 & 3.45 & 0.22 \\
    Cloud B & 2.64 & 3.10 & 0.07 (0.14) & 2.95 & 3.41 & 0.21 \\
    Cloud C & 2.72 & 3.14 & 0.08 (0.16) & 3.07 & 3.49 & 0.19 \\
    Cloud D & 2.68 & 3.42 & 0.09 (0.18) & 2.98 & 3.79 & 0.18 \\
    Cloud E & 2.78 & 3.29 & 0.08 (0.17) & 3.11 & 3.70 & 0.20 \\
    Cloud F & 2.65 & 3.47 & 0.09 (0.19) & 3.07 & 3.92 & 0.16 \\
    Cloud G & 2.69 & 3.12 & 0.07 (0.15) & 3.05 & 3.50 & 0.20 \\
    Cloud H & 2.67 & 2.98 & 0.08 (0.16) & 3.00 & 3.36 & 0.20 \\
    Cloud I & 2.61 & 3.10 & 0.07 (0.16) & 2.98 & 3.38 & 0.21 \\
    Cloud J & 2.66 & 3.57 & 0.08 (0.19) & 2.94 & 4.08 & 0.22 \\
    
    \hline\hline
    
    \end{tabular}

\end{table}
\section{Results}\label{sec:results}

The ALMA 3\,mm dust continuum emission maps, and maps of the integrated intensity of \ntwohoz\ are presented in Figure\,\ref{fig:mom_maps} (second and third column, respectively). To produce the integrated intensity maps, we use cubes with a rest frequency centred on the isolated hyperfine component of \ntwohoz, and integrate emission between the systemic velocities shown in Table\,\ref{tab:cloudprops}. In Figure\,\ref{fig:mom_maps} (first column) we also show a three colour {\it Spitzer} GLIMPSE survey map, where the IRDCs can be seen as dark extinction features \citep{carey_2009}, and near- and mid-infrared extinction derived mass surface density maps (see fourth column; \citealp{kainulainen_2013}). We find that a complex extended filamentary structure is present within the \ntwohoz\ maps, which appears to be broadly similar to the structure for each cloud as seen in the infrared extinction observations. The continuum emission observations, on the other hand, appear much less extended. A comparison to the infrared images shows that the continuum maps recover only the extinction peaks, or infrared point sources (seen as green sources in the three-colour image, and as holes in the mass surface density maps). 

\begin{figure*}
\centering
    \includegraphics[width=1\textwidth]{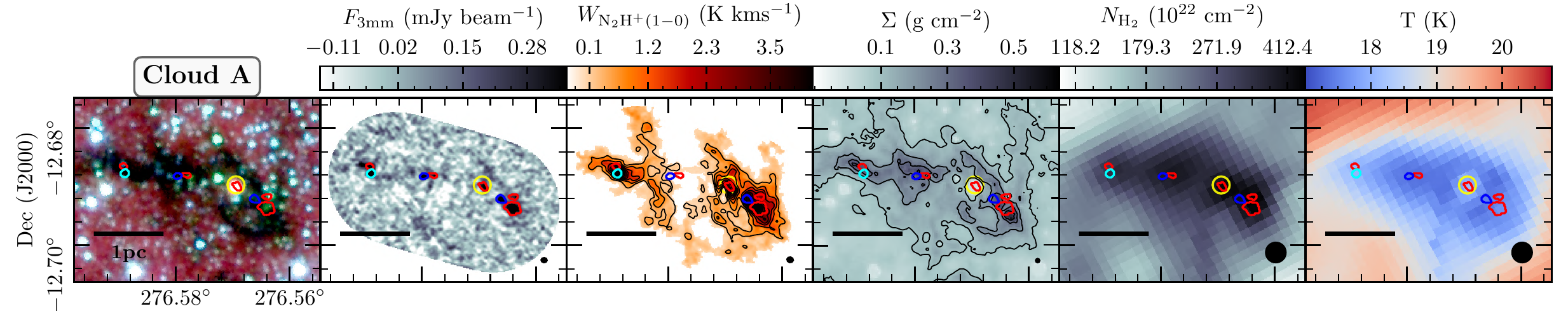}
    \includegraphics[width=1\textwidth]{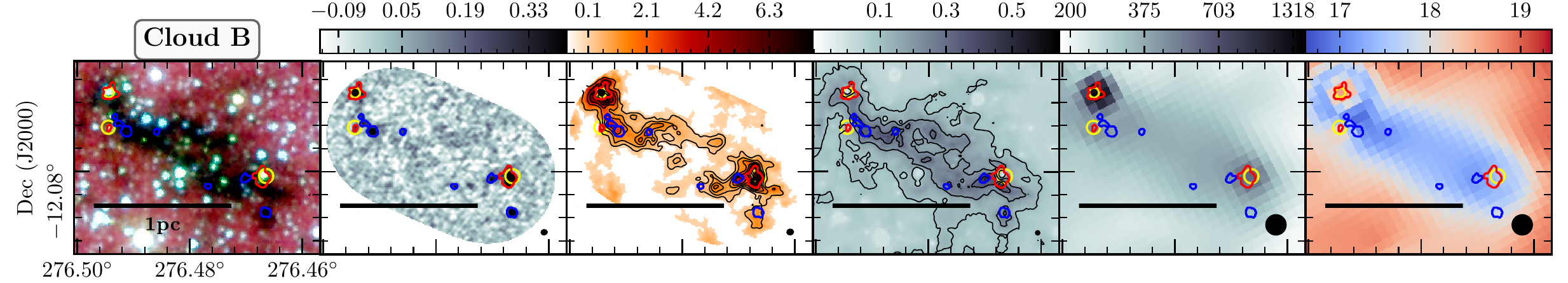}
    \includegraphics[width=1\textwidth]{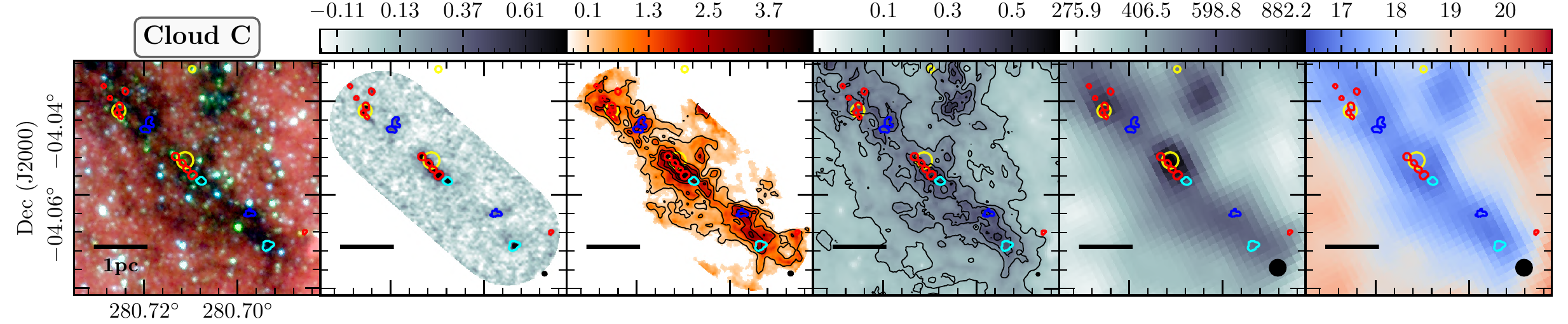}
    \includegraphics[width=1\textwidth]{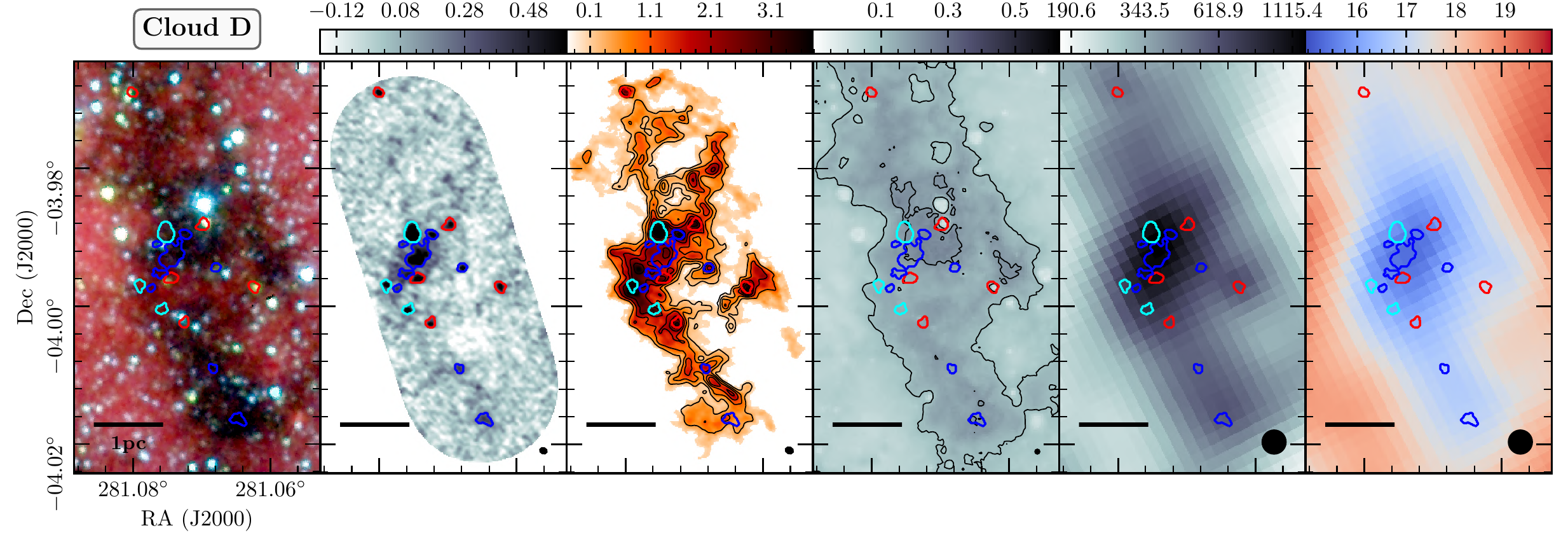}
    
    	\caption{(first column) Three colour images from the {\it Spitzer} GLIMPSE survey \citep{churchwell_2009}. (second) The ALMA 3\,mm dust continuum (without the primary beam correction), and (third) \ntwoh\,(J,F${_1}$,F = 1,0,1 $\rightarrow$ 0,1,2) integrated intensity maps. (fourth) Combined near- and mid-infrared extinction derived mass surface density maps \citep{kainulainen_2013}. (fifth) Far-infrared {\it Herschel} derived column density, and (sixth) dust temperature maps \citep{marsh_2016, marsh_2017}. \note{The yellow circles overlaid on each panel show the positions and sizes of 70\,\micron\ emission point sources \citep{molinari_2016,marton17}.} The coloured contours overlaid on each panel show boundaries of all the ``leaves'' (or cores) identified using the dendrogram analysis on the dust continuum maps (see section\,\ref{sec:coreident}). \note{The red and blue contours indicate the cores that have been classified as star-forming and quiescent, respectively (section\,\ref{sec:coreident}).} The cyan contours indicate the cores that have been identified as being quiescent, high-mass and without further fragmentation (see section\,\ref{sec:hmcores}). Shown in the lower left of the ALMA observation panels is the beam size, and in the lower left of the three colour image (first panel) is a scale bar adjusted for the distance of each cloud (see Table\,\ref{tab:cloudprops}).} 
    	
    	\label{fig:mom_maps}
\end{figure*}

\begin{figure*}
\centering
    \includegraphics[width=1\textwidth]{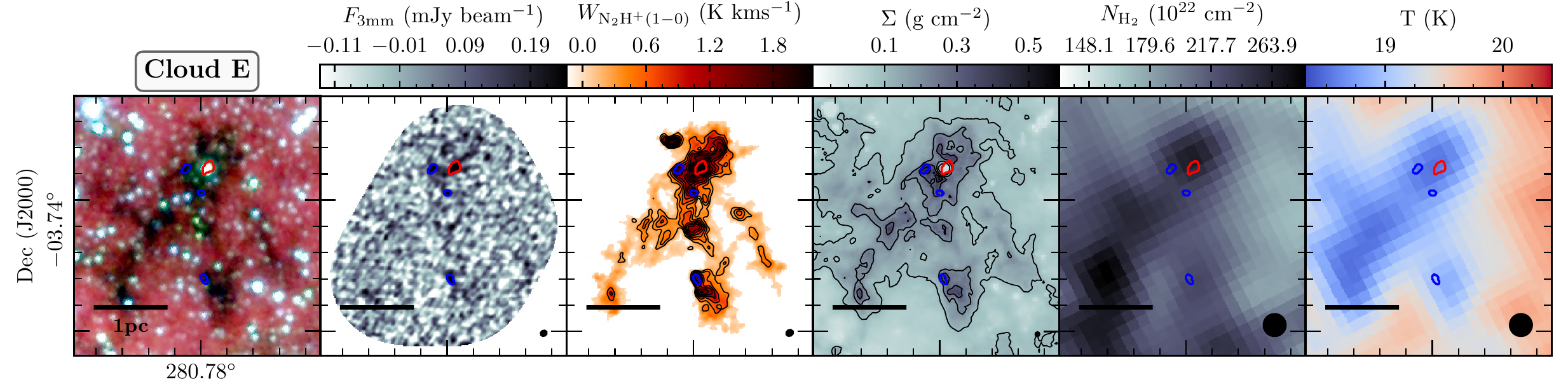}
    \includegraphics[width=1\textwidth]{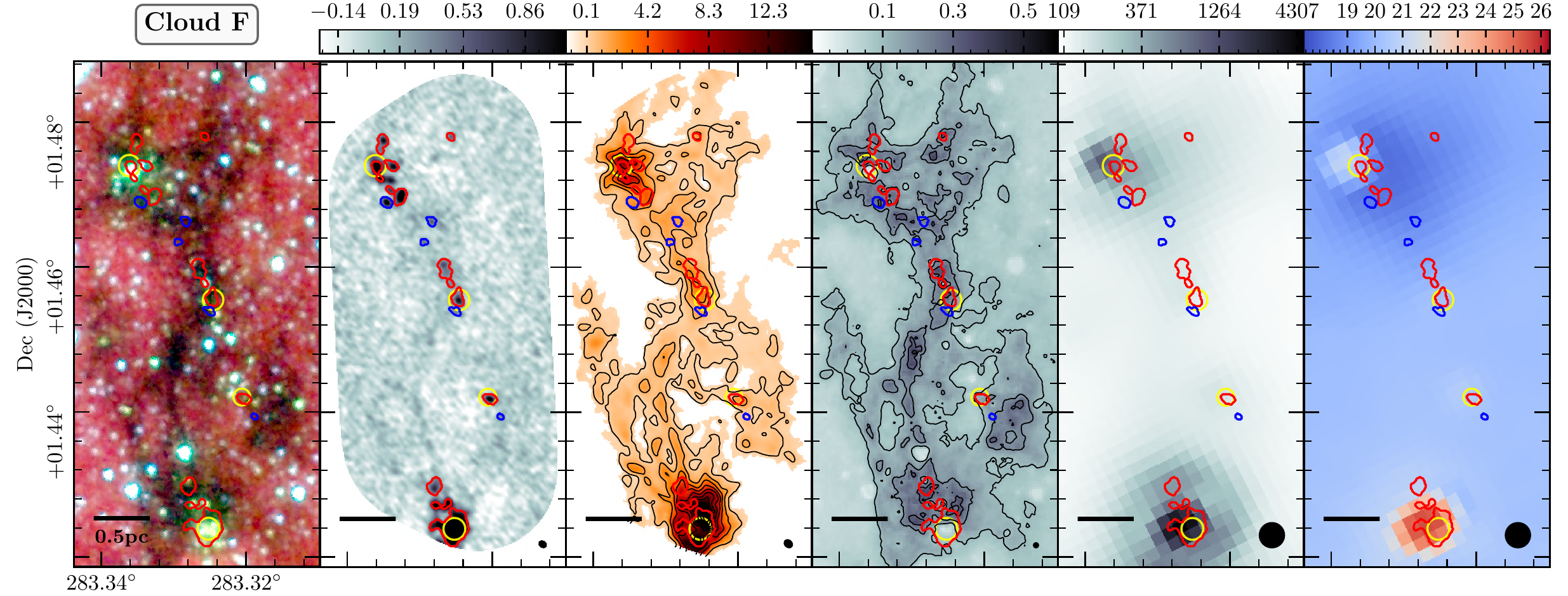}
    \includegraphics[width=1\textwidth]{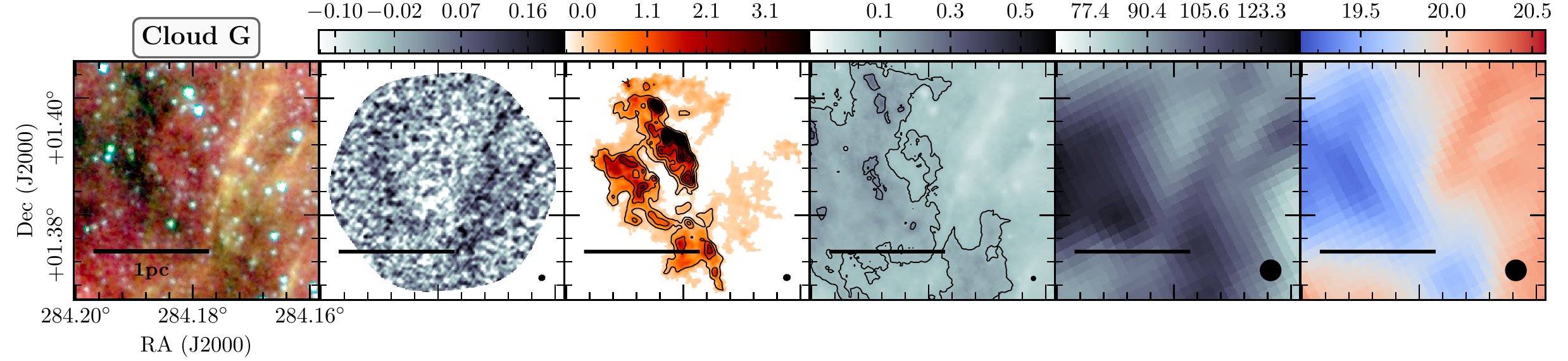}
    	\contcaption{} 
    	\label{}
\end{figure*}

\begin{figure*}
\centering
    \includegraphics[width=1\textwidth]{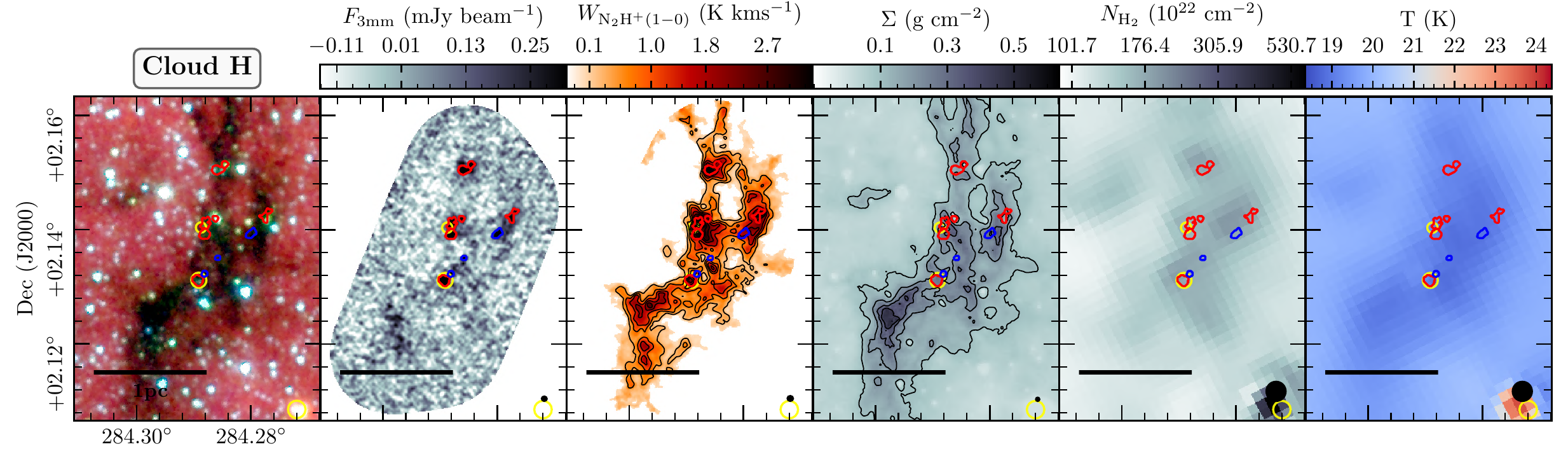}
    \includegraphics[width=1\textwidth]{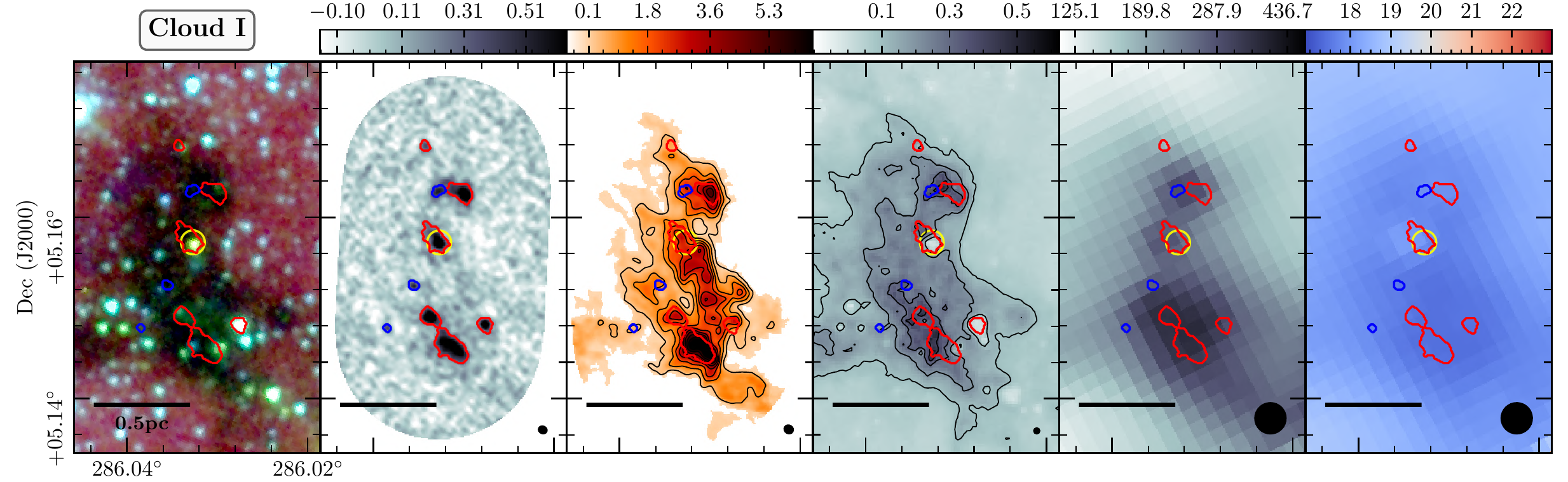}
    \includegraphics[width=1\textwidth]{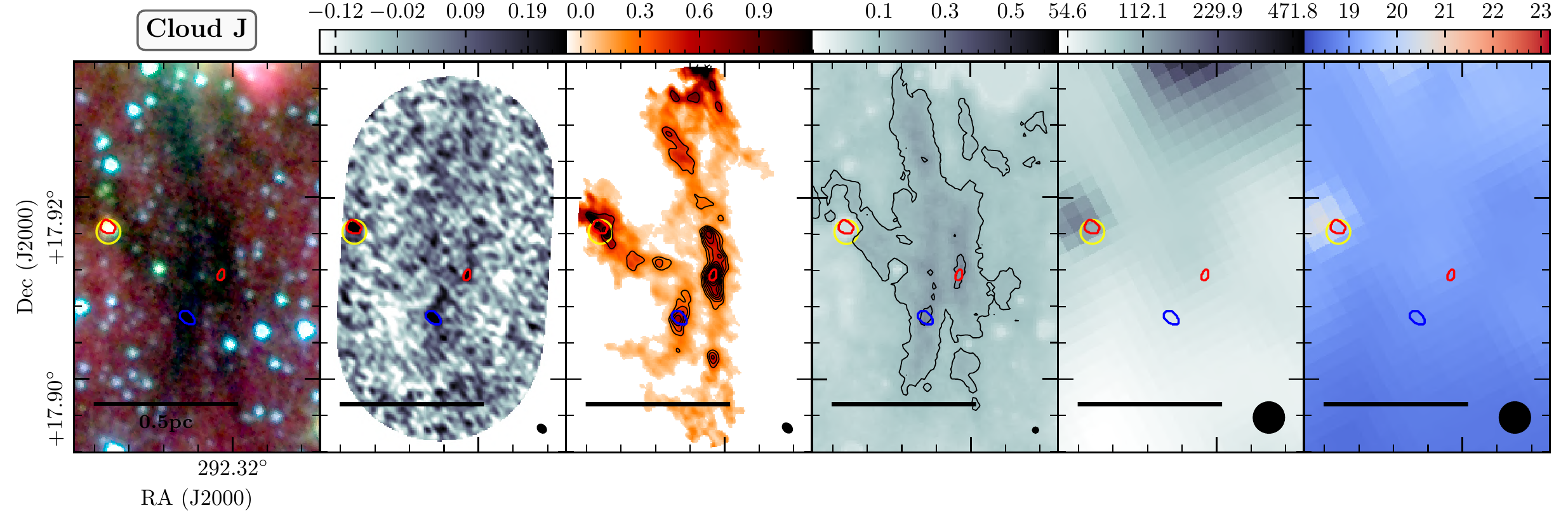}
    	\contcaption{} 
    	\label{}
\end{figure*}

\subsection{Core identification}\label{sec:coreident}

Despite the relatively simple morphology of the continuum emission, we characterise the structures present across the sample using a dendrogram analysis \citep{rosolowsky_2008}. The use of a structure-finding algorithm, as opposed to by-eye identification, is preferred to give reproducibility and allow a systematic comparison across the sample. Dendrogram analysis, in particular, was chosen to allow a more direct comparison to other works which cross-over with our cloud sample \citep{henshaw_2016c,henshaw_2016d,liu_2018}. 

We run the dendrogram analysis using the combined 12\,m and 7\,m array continuum maps that have not been corrected for the primary beam response. These maps have a flat noise profile, and, therefore, are preferred over the primary beam corrected maps, as the initial determination of the dendrogram structure relies on the constant noise threshold calculated for each cloud (see Table\,\ref{tab:obsprops}). We found that the use of the primary beam corrected maps typically caused the dendrogram algorithm to identify noise features towards the edge of the mapped region as significant structures. We tested a range of input parameters for the dendrogram analysis of the non-primary beam corrected maps, and found that a set of parameters similar to \citet{liu_2018} produced the structure that best resembled what would be identified through manual inspection of the continuum maps. Along with allowing a more direct comparison to the \citet{liu_2018} results, this parameter set has the benefit of being well tested for several additional data sets covering the same sources observed in this work (e.g. \citealp{henshaw_2016c, cheng_2018}). The set of parameters that are used for the determination of the dendrogram structure used throughout this work is: {\sc min\_value} = 3\,$\sigma$=0.24\,mJy\,beam$^{-1}$ (the minimum intensity considered in the analysis); {\sc min\_delta} = 1\,$\sigma$ (the minimum spacing between isocontours); {\sc min\_pix} = 0.5\,beam area $\sim$ 18 pixels (the minimum number of pixels contained within a structure).We tested these parameters, and found that minor changes were required to achieve dendrogram hierarchies that include all structures identified from manual inspection of the continuum maps. It is worth keeping in mind that the choice of any parameter set in an automated structure identification algorithm only allows for a reproducible structure, and, ultimately, the user must check the results for inherently complex datasets. 


Overlaid as coloured contours on the maps shown in Figure\,\ref{fig:mom_maps} are the leaves identified from the dendrogram analysis. The leaves represent the highest level (smallest) structures in the dendrogram analysis, which we refer to as ``cores'' within this work. The positions and effective radii ($R_\mathrm{eff} = \sqrt{A/\pi}$, where $A$ is the area enclosed within the dendrogram boundary) of the 96 identified cores can be found in Table\,\ref{tab:core_obsprops}. We find that around 5 to 15 cores are present within each cloud, with the exception of Cloud G where no cores have been identified. This was likely because the mapped region of Cloud G is, by design, focused on the eastern shocked region explored by \citet{cosentino_2018, cosentino_2019}, and not the main dust extinction/continuum feature(s) previously identified within this cloud (e.g. \citealp{rathborne_2006, butler_2012, kainulainen_2013}).

Whereas the dendrogram structure itself has been calculated using the non-primary beam corrected continuum maps, all fluxes quoted in this work have been corrected for the primary beam response. Moreover, when determining the total primary beam corrected flux for each core, we consider that these are not isolated structures, but rather that they are sitting within the complex three-dimensional geometry of their clouds. They, therefore, may have some fore- and back-ground flux contribution from their host environment \citep{rosolowsky_2008}. We, therefore, follow two methods to determine the total flux within a given core boundary. The first assumes that there is no background contribution of the flux, that is that all of the flux within the leaf boundary is attributed to that structure ($S_\nu$). The second approach is to assume that a core is superimposed on top of the background flux level, which needs to be subtracted to get the ``background subtracted flux'' ($S^\mathrm{b}_\nu$). In practice, we take the structure that is directly below a given leaf in the dendrogram hierarchy as its background level (i.e. the branch where the leaf is located), and determine the background subtracted flux as the remaining flux after having subtracted the contribution of this lower level structure from the leaf (e.g. \citealp{pineda_2015, henshaw_2016}). All properties determined within this work using this background-subtracted flux will be denoted by a superscript ``b'' (e.g. $M^\mathrm{b}$ is the background-subtracted mass). We find that the total flux contribution from the leaves after subtraction of background emission is $S^\mathrm{b}_\nu/S_\nu = 0.29_{-0.11}^{+0.23}$.\footnote{For all statistics we present within this section, we show the median (50 percentile) of the sample distribution, and one standard deviation around this value (15.9 and 84.1 percentiles).} These flux values are summarised in Table\,\ref{tab:core_obsprops}.

\note{One of the primary aims of this work is to identify and study the earliest stages of high-mass star formation. Therefore, finally, we determine if the cores (i.e. dendrogram leaves) contain any near- or mid-infrared emission, which could be suggestive of them being at a later evolutionary stage and potentially harbouring active star formation (e.g. \citealp{ragan_2012, rigby21}). To do so, we compare the cores to the {\it Spitzer} 3.6\micron\ (blue), 4.5\micron\ (green) and 8\micron\ (red) three-colour images \citep{churchwell_2009}, as shown in Figure\,\ref{fig:mom_maps} (first column). We then visually determined if there is an infrared point source within each of the core boundaries as defined by the dendrogram contours. These associations are then cross-referenced with the {\it Herschel} 70\micron\ emission maps and point source catalogue, also overlaid as circles on Figure\,\ref{fig:mom_maps} \citep{molinari_2016, marton17}. We find that the majority of 70\micron\ point sources have corresponding {\it Spitzer} emission ($\sim$90 per cent), yet many of the {\it Spitzer} emission sources do not have 70\micron\ emission ($\sim$60 per cent). Sources containing either or both {\it Spitzer} emission or a 70\micron\ point source are labelled as star-forming in our catalogue. This combination of infrared emission allows for a rigorous detection of both early and later stages of embedded star formation across our core sample. The star-forming state of each core is provided in Table\,\ref{tab:core_obsprops}. In Figure\,\ref{fig:mom_maps}, quiescent cores are represented by the blue and cyan contours (see section\,\ref{sec:hmcores} for a discussion of the cyan contours), and star-forming cores are represented by red contours on Figure\,\ref{fig:mom_maps}.}

\subsection{Physical properties}\label{sec:physprops}

\begin{figure*}
\centering
    \includegraphics[width=1\textwidth]{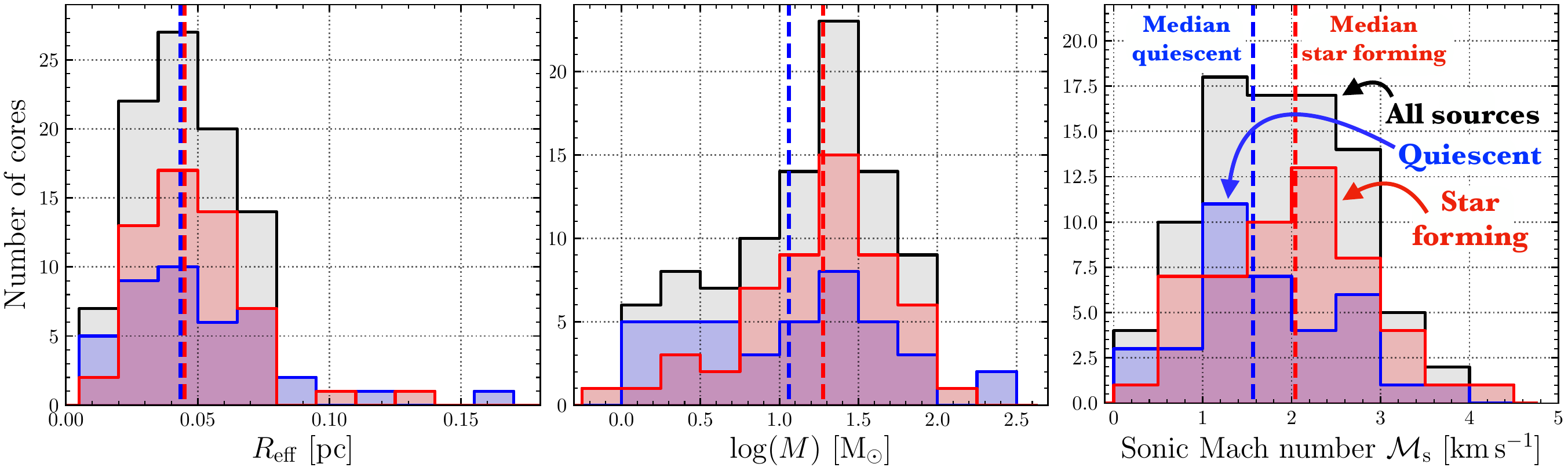}
    	\caption{Distribution of the properties determined across the core sample (section\,\ref{sec:dynprops}). Shown in panels from left to right is the effective radius ($R_\mathrm{eff}$), the (logarithm) mass ($M$), and sonic Mach number ($\mathcal M_\mathrm{s}\,=\,\sigma _\mathrm{NT} / \sigma _\mathrm{T}$) calculated across the core sample (section\,\ref{sec:dynprops}). Shown as black, red and blue stepped profiles are histograms for the whole sample, the star-forming cores and quiescent (non-star forming) cores, respectively. The red and blue vertical dashed lines show the median values for the star-forming cores and quiescent cores.}
    	\label{fig:hist_mach}
\end{figure*}

In this section, we determine the physical properties of the core catalogue identified using the dendrogram analysis (section\,\ref{sec:coreident}). Firstly, we determine the projected size (or effective radius) and masses of each core. We find an angular size distribution of $R_\mathrm{eff}=2.4_{-0.8}^{+0.8}$ arcsec, which highlights that the majority of the cores have sizes only marginally larger than the mean beam radius of $\sim\,1.5-2$ arcsec, and, therefore, are not fully resolved (see Table\,\ref{tab:obsprops}). This angular size distribution corresponds to a projected size distribution of $R_\mathrm{eff}=0.04_{-0.02}^{+0.02}$\,pc (or $9210_{-4197}^{+4844}$\,AU), when accounting for the source distances given in Table\,\ref{tab:cloudprops}. \note{The uncertainty in $R_\mathrm{eff}$ from observational errors (e.g. pointing) is negligible compared to the uncertainty introduced from the kinematic distance. \citet{simon_2006b} estimated that the kinematic distances are typically accurate to $\sim$\,15\,per cent, which we adopt for our uncertainty on the radius.}

To calculate the mass of each core, we use the integrated flux ($S_\nu$; see Table\,\ref{tab:core_obsprops}) following, 
\begin{equation}
    M = \frac{d^{2} S_\nu R_\mathrm{gd}}{\kappa_\nu B_\nu(T_\mathrm{dust})}, 
\label{equ:mass}
\end{equation}
where $d$ is the source distance (see Table\,\ref{tab:cloudprops}), $R_\mathrm{gd} = 141$ is the total (gas plus dust)-to-(refractory-component-)dust-mass ratio (assuming a typical interstellar composition of H, He, and metals; \citealp{draine11}),\footnote{Here we adopt a higher than typically assumed value for the dust-to-gas ratio, which determined from $M_\mathrm{H}$/$M_\mathrm{dust}=101$ (see Table\,21.3 of \citealp{draine11}), or $M_\mathrm{total}$/$M_\mathrm{dust}=1.4\times101=141$
(see Table 1.4 of \citealp{draine11} for $M_\mathrm{total}$/$M_\mathrm{H}=1.4$).} $B_\nu(T_\mathrm{dust})$ is the Planck function for a dust temperature, $T_\mathrm{dust}$, at a representative frequency of $\nu$\,=\,90.664\,GHz, and $\kappa_\nu = \,\kappa_0 \left( \nu / \nu_{0} \right)^\beta \approx 0.175$\,cm$^2$g$^{-1}$, when assuming $\nu_{0}$\,=\,230\,GHz, $\beta = 1.75$ \citep{battersby_2011} and $\kappa_0$\,=\,0.899\,cm$^{2}$\,g$^{-1}$ (\citealp{ossenkopf_1994} result for an MRN size distribution with thin ice mantles after $10^{5}$\,yr of coagulation at a density of 10$^{6}$\,\cmcb).

To obtain an estimate of the dust temperature towards each of the cores, we make use of spectral energy distribution (SED) fits to the far-infrared dust continuum observed with {\it Herschel} \citep[Hi-Gal;][]{molinari_2016}. We use the mean line-of-sight results from the PPMAP project shown in Figure\,\ref{fig:mom_maps} \citep{marsh_2016, marsh_2017}. To check the fidelity of the PPMAP column density and temperature maps, where possible, we compare to the corresponding maps from \citet{nguyen_2011}, \citet{lim16}, \citet{zhang_2017}, and \citet{soam_2019}. These authors independently produced column density and temperature maps following a more conventional far-infrared SED fitting routine, whilst accounting for the background in {\it Herschel} observations. We find broadly comparable values of both the column density and temperature within these conventional maps and the PPMAP maps, albeit the dust temperature determined with PPMAP appears to be a few degrees higher towards the cloud centres (also see \citealp{marsh_2017} for a similar comparison). As this does not significantly affect the results of this paper, and because the PPMAP dataset is the only available consistent set of maps for the full IRDC sample, for consistency we choose to use the PPMAP results (additional sources of uncertainty on e.g. the dust temperature measurements are discussed throughout this work). 

We find dust median temperatures of $T_\mathrm{dust}=17.8_{-0.9}^{+1.4}$\,K. These dust temperatures for each core are then used to calculate a median mass of $M=18.4_{-15.0}^{+32.5}$\,\sol\ across the core sample ($M^\mathrm{b}=4.6_{-3.7}^{+16.8}$\,\sol).\footnote{We also determine the masses for a fixed temperature of 18\,K, which approximately corresponds to the mean dust temperature determined across the cloud sample.} We assumed a typically $\sim$\,10\,per cent in the absolute flux scale of the ALMA observations, and, following \citet{Sanhueza2017}, we assume an uncertainty of $\sim$\,30 per cent dust opacity. These uncertainties in the dust opacity, dust emission fluxes, and the distance propagate to give an uncertainty of $\sim$\,50 per cent in masses. The histogram distribution of the size and mass of the star-forming and quiescent core samples is shown in Figure\,\ref{fig:hist_mach}.

It is worth noting that the dust temperatures measured here are averages sampled by the beam of the PPMAP maps ($\sim$12\arcsec; shown as a black circle in the rightmost panel of Figure\,\ref{fig:mom_maps}), and the beam of the ALMA observations used in this work is significantly smaller ($\sim$\,3\arcsec; see Table\,\ref{tab:obsprops}). Therefore, smaller-scale temperature variations due to e.g. cold cores or embedded stellar objects could cause us to over- or under-estimate the temperature, respectively (e.g. see \citealp{ragan_2011, dirienzo_2015, sokolov_2017}). 

\note{To quantify the extent that the cores containing embedded protostars may have underestimated temperatures in the PPMAP maps, we follow a procedure outlined in \citet{peretto20} for determining the temperature at higher angular resolution ($\sim$5\arcsec). For this estimate, these authors require that the cores have an identified 70\micron\ point source (the 70\micron\ flux given in Table\,\ref{tab:core_obsprops}). The 70\micron\ flux is then converted to a bolometric flux following the relation from \citet{elia17}, which is thought to apply to high-mass star forming regions. Following flux conservation, this luminosity can then be converted to a temperature profile \citep{terebey93}. Using the same fiducial parameters as \citet{peretto20}, we determine the mass-averaged temperature for each of our 70\micron\ cores. We find that temperature estimates from the 70\micron\ emission are systematically higher than those within the PPMAP. The mean difference in temperature is a factor of 50 per cent, which on average corresponds to 30 per cent lower 70\micron\ mass estimates, $M_\mathrm{70\mu m}$ (see Table\,\ref{tab:core_physprops}). These differences are included as additional uncertainties for our star-forming sample, which propagate to a $\sim$\,60 per cent in their mass estimates from PPMAP that are adopted throughout this work.}

We also obtain masses from the extinction derived mass surface density maps of the clouds (\citealp{kainulainen_2013}; as shown in Figure\,\ref{fig:mom_maps}). These maps have a comparable angular resolution to our ALMA observations, and, therefore, should serve as an independent measure of the core masses. Extracting the extinction total mass within the core boundaries gives a median value of $M_\mathrm{ext}=6.3_{-4.7}^{+11.0}$\,\sol, which is a significant fraction below the continuum derived masses: $M_\mathrm{ext}/M=0.5_{-0.3}^{+0.4}$. This offset was also noted by \citet{henshaw_2016c}, who compared 3\,mm dust continuum and extinction derived masses within the northern portion of Cloud H, not included in these observations. We assess if this could be a product of the artificially lower values within the mass surface density maps, where infrared point source emission inhibits an accurate extinction measurement (e.g \citealp{kainulainen_2009, butler_2009, butler_2012, kainulainen_2013}).\footnote{We note that the mass surface density maps contain negative values towards the brightest infrared point sources. As a result of this, there are two cores associated with infrared point sources that have negative masses, which we remove from our sample when calculating the $M_\mathrm{ext}$ statistics.} These can be seen by comparing the infrared three colour image and mass surface density maps within Figure\,\ref{fig:mom_maps}. 

\begin{figure}
\centering
    \includegraphics[width=1\columnwidth]{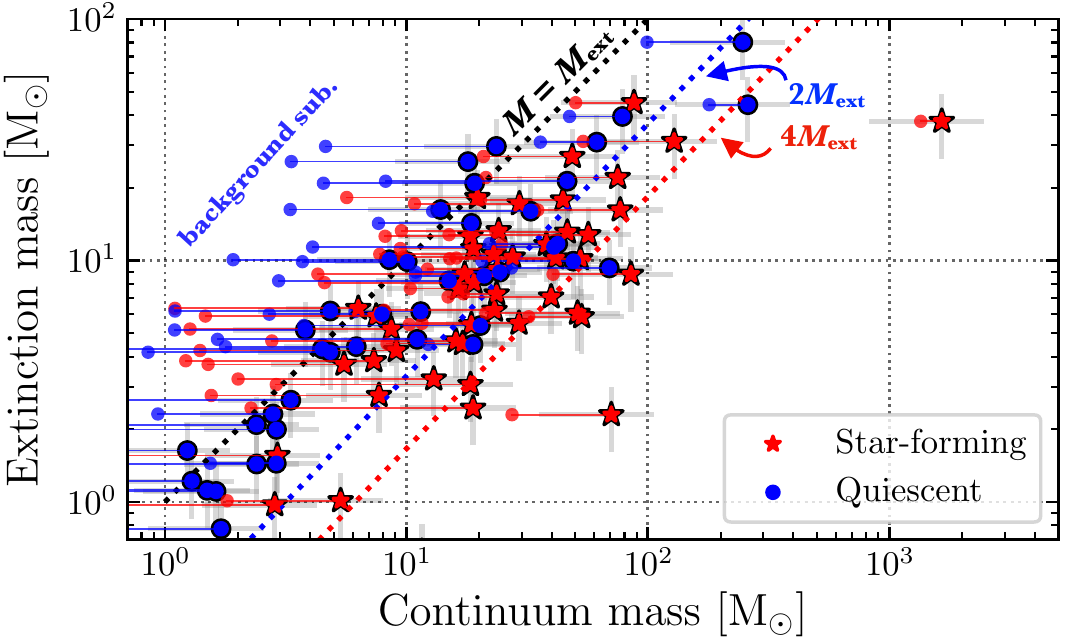}
    	\caption{A comparison between the core masses determined from the 3\,mm ALMA continuum and combined near- and mid-infrared extinction (section\,\ref{sec:physprops}). The cores classified as actively star-forming or quiescent are shown in red and blue, respectively. Shown as points with and without a black-outline are the mass estimates determined without and with background-subtraction (see Table\,\ref{tab:core_physprops}). \note{We show the uncertainty ranges of $\sim$\,50 and $\sim$\,60\,per cent for star forming continuum mass estimates, respectively, and $\sim$\,30\,per cent for the extinction masses estimates \citep{butler_2009,butler_2012,kainulainen_2013}.} The diagonal dotted lines show the continuum and extinction determined masses are equal ($M=M_\mathrm{ext}$), and the continuum determined mass is a factor of two and four higher ($M=2M_\mathrm{ext}$, $M=4M_\mathrm{ext}$).} 
    	\label{fig:extcomp}
\end{figure}

Figure\,\ref{fig:extcomp} shows the core masses determined from the near- and mid-infrared extinction as a function of the core masses determined from the 3\,mm ALMA continuum, where cores with and without signs of star formation are shown in red and blue, respectively. Here we see that the cores are systematically offset towards higher continuum determined masses by around a factor of two to three (compare to black dashed lines). Moreover, we see that the ratio of the extinction to continuum mass estimates ($M_\mathrm{ext}/M$) with signs of star formation are systematically lower than those without star formation. We calculate that the median ratio with and without star formation is $M_\mathrm{ext}/M=0.29$, $M_\mathrm{ext}/M=0.68$, respectively. It is then reasonable to suggest that the larger difference for the cores with star formation is primarily a result of the extinction maps having lower values of the mass surface density towards bright infrared point sources (e.g \citealp{kainulainen_2009, butler_2009, butler_2012, kainulainen_2013,butler14}). However, the remaining factor of two difference between the continuum and extinction determined masses for the non-star-forming sources is not clear (also see \citealp{henshaw_2016,liu_2018,kong18}). 

In Figure\,\ref{fig:extcomp}, we also show the background-subtracted continuum masses for reference. However, we note that the extinction derived masses have not been background subtracted. Hence, this is not direct comparison, and the inclusion of background subtraction can not stand as an explanation for the remaining difference. Varying the assumed dust properties within reasonable limits can not account for the factor of four difference in mass estimates for the quiescent sources. \note{In addition, we note that the masses of the star forming sources determined using the 70\,\micron\ temperature estimate are still systematically higher than the extinction based estimate ($M_\mathrm{ext}/M_\mathrm{70\mu m}=0.35$).} We then attribute this to the variety of systematic uncertainties inherent in the extinction mapping technique, which include foreground corrections that affect lower column density regions, and saturation effects at high optical depths that cause the mass in high mass surface density regions to be underestimated \citep{butler_2009, butler_2012, kainulainen_2013, butler14}.

The molecular hydrogen number density of each core is determined using, 
\begin{equation}
    n_\mathrm{H_2} = \frac{M}{\frac{4}{3} \pi R_\mathrm{eff}^3 \mu_\mathrm{H_2} m_\mathrm{H}},
\end{equation}
where $\mu_\mathrm{H_2} = 2.8$ is the mean molecular weight per hydrogen molecule \citep{kauffmann_2008}, and $m_\mathrm{H}$ is the mass of a hydrogen atom. We find molecular hydrogen number densities across the sample of $n_\mathrm{H_2} = 6.9_{-3.2}^{+5.6}\times10^5$\,cm$^{-3}$, or when using the background subtracted mass $n^\mathrm{b}_\mathrm{H_2} = 2.0_{-1.2}^{+2.3}\times10^5$\,cm$^{-3}$. The corresponding local free-fall time is calculated as, 
\begin{equation}
    t_\mathrm{ff} = \left( \frac{\pi^2 R_\mathrm{eff}^3}{8 GM } \right ) ^{0.5} = \left( \frac{3 \pi}{32 G \mu_\mathrm{H_2} m_\mathrm{H} n_\mathrm{H_2}} \right ) ^{0.5},
\end{equation}
where $G$ is the gravitational constant. We find local free-fall times across the sample of $t_\mathrm{ff} =3.7_{-1.0}^{+1.4}\times10^4$\,years. The number densities and free-fall times for each core are given in Table\,\ref{tab:core_physprops}.

\subsection{Dynamical properties}\label{sec:dynprops}

We now determine the core dynamical properties, which will be used later to assess the stability of the cores. We use the (1-0) transition of \ntwoh\ to determine the line-of-sight velocity dispersion for each of the cores. To do so, we extract the average spectra from within the leaf contours.\footnote{The spectra have been adjusted to a rest frequency of 93176.2522\,MHz, corresponding to the isolated hyperfine component J,F$_1$,F = 1,0,1 $\rightarrow$ 0,1,2, from the original rest frequency of 93176.7637\,MHz that corresponds to the brightest hyperfine component J,F$_1$,F = 1,2,3 $\rightarrow$ 0,1,2 \citep{caselli_1995a, pagani_2009}.} We fit each of the spectra using the stand-alone fitter functionality of {\sc ScousePy} \citep{henshaw_2019}. This procedure uses a technique called derivative spectroscopy to provide an estimate of the number of emission features within each spectrum \citep{lindner_2015,riener_2019}. Briefly, this technique involves smoothing the spectrum using a Gaussian kernel. Features such as maxima, minima, and inflection points detected within the derivatives of this smoothed spectrum can be then used to determine estimates of the peak amplitude, centroid velocity, and width of each emission feature within a spectrum. {\sc ScousePy} feeds these values as free-parameter guides to {\sc PySpecKit} \citep{ginsburg_2011}, which performs the fitting. The size of the smoothing kernel is provided as an input parameter and is somewhat data dependent \citep[][use machine learning to determine this parameter]{riener_2019}, however, we set a value of either 3, 4, 5 or 6 for 1, 1, 59 and 35 cases, respectively. {\sc ScousePy} also allows for manual intervention. Manual fitting was performed in cases where the width of the main group of hyperfine components is such that the emission encroached within our fixed window around the isolated component, leading to the identification of peaks that we do not want to fit. We find that 47 out of the 96 cores required multiple Gaussian components to accurately reproduce the observed spectra. The majority of these (38) required only an additional component, but we note that 9 cases had particularly complex spectra that required three components. In these cases with multiple velocity components, we assign the component with the largest integrated intensity to the core. 

\begin{figure}
\centering
    \includegraphics[width=1\columnwidth]{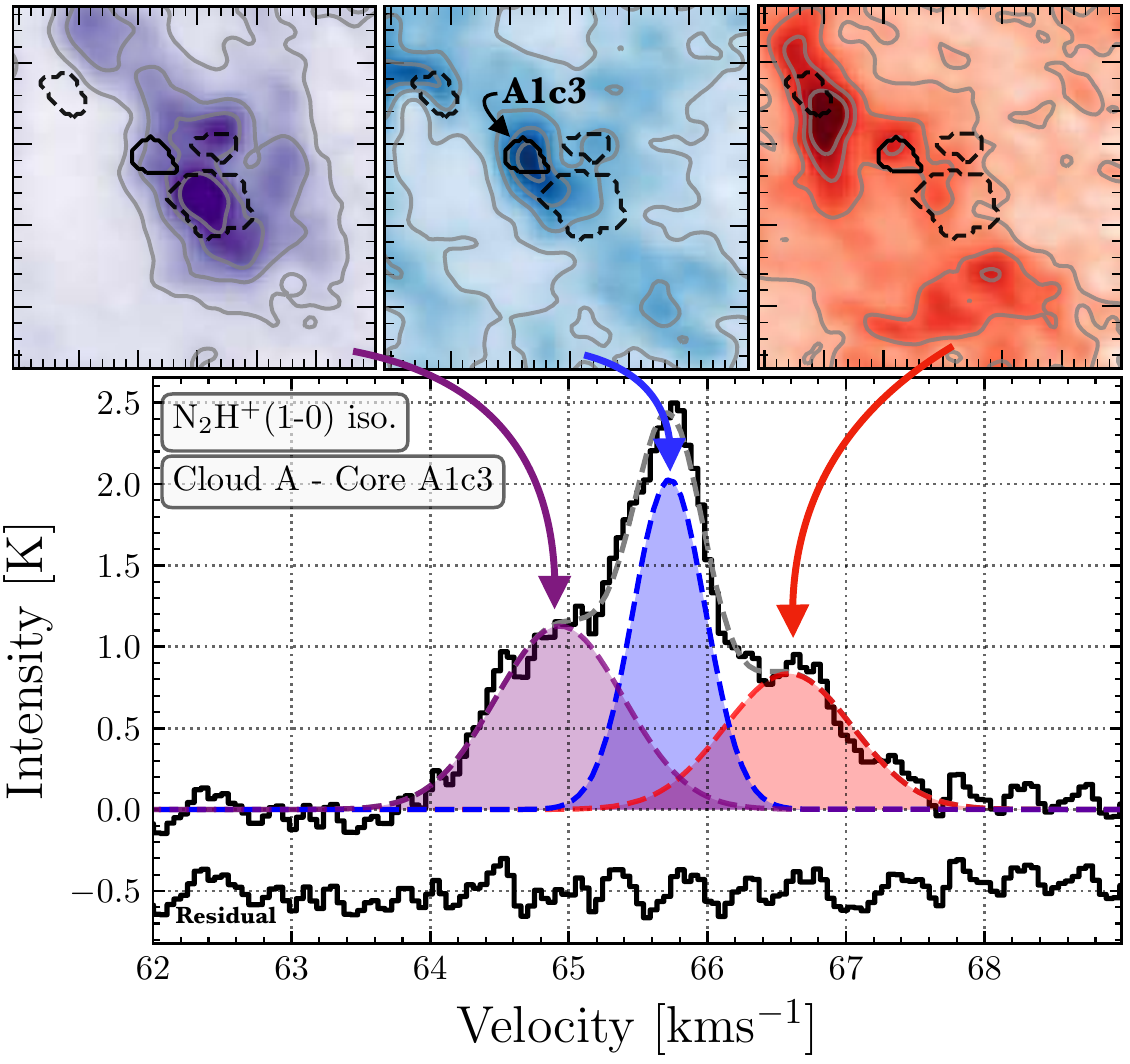}
    	\caption{Gaussian decomposition of the velocity components towards core A1c3 (section\,\ref{sec:dynprops}). The spectrum of the isolated hyperfine component of \ntwohoz\ is shown as a black solid line, whilst the three velocity components identified by the {\sc ScousePy} routine are shown as the dashed and filled coloured lines. The total fitted spectrum is shown as a grey dashed line, and the residual of observed spectrum minus the total fitted line is shown as a solid black line centred at -0.5\,K. Shown in the above panels are maps towards Core A1c3 of the \ntwohoz\ intensity integrated over the approximate velocity range covered by each component. These maps have been overlaid with grey contours also showing the \ntwohoz\ intensity, in levels that have been chosen to best highlight the emission morphology. Shown as a solid black contour is the boundary of Core A1c3, and the dashed black contours show the remaining cores within this region (also see Figure\,\ref{fig:mom_maps}). This Figure highlights that the brightest velocity component has the best spatial correspondence to the continuum peak.}
    	\label{fig:spec}
\end{figure}

An example of a complex spectral profile observed towards core A1c3 is shown in Figure\,\ref{fig:spec}. This spectrum has been decomposed into 3 velocity components by {\sc ScousePy}, which are shown as the purple, blue and red dashed, filled Gaussian profiles. The sum of these components is shown as a dashed grey profile, and the residual of the observed spectrum minus the total profile is shown as a solid black centred on -0.5\,K. This residual profile contains variations at the level of the noise, which validates the choice of fitting parameters to accurately reconstruct the observed profile. Also shown in Figure\,\ref{fig:spec} are maps of the \ntwohoz\ intensity integrated over the approximate velocity range covered by each component. These are shown with colour-scales that match the colours used to show the Gaussian profiles. Overlaid as dashed black contours on these maps are the cores, and the solid black contour corresponds to core A1c3; over which the spectrum shown in the main panel has been taken. These panels clearly show that, although several velocity components overlap at the region of core A1c3, only the brightest component (shown here in blue) has a spatial morphology that peaks at the position of the core A1c3. The purple and red components appear to peak more towards cores A1c1/2 and A1c4, respectively. This then validates the assumption that the brightest velocity component observed in \ntwohoz\ is associated with the continuum core. We perform a similar manual check of the \ntwohoz\ emission morphology towards each core and find that this assumption also holds in the vast majority of cases. For the 2$-$5 potential exceptions (i.e. $\sim$5\,per cent of the overall sample), it is, however, not definitively clear if the choice of a lower brightness velocity component would be preferable. These will be investigated further in a future work in this series, in which we will conduct a more comprehensive pixel-by-pixel decomposition of the velocity structure within each cloud (Henshaw et al. in prep.).

We correct the observed velocity dispersion, $\sigma_\mathrm{obs}$, for the minor contribution of the velocity resolution,   
\begin{equation}
    \sigma_\mathrm{v}^2 = \sigma_\mathrm{obs} ^2 - \frac{\Delta \mathrm{v}_\mathrm{res}^2} {8 \,\mathrm{ln} 2},
\label{equ:velocity_disp}
\end{equation}
where $\mathrm{v}_\mathrm{res}\sim0.1$\,\kms\ is the velocity resolution of the observations (see section\,\ref{sec:observations}). Moreover, we determine the contribution of the non-thermal motions to the velocity dispersion. This can be calculated as \citep{fuller_1992, henshaw_2016c},
\begin{equation}
    \sigma_\mathrm{NT}^2 = \sigma_\mathrm{v}^{2} - \sigma_\mathrm{T}^{2} = \sigma_\mathrm{v}^{2} - k_\mathrm{B}T_\mathrm{kin} \left( \frac{1}{\bar{m}} - \frac{1}{m_\mathrm{obs}} \right),
\end{equation}
\noindent where $\sigma_\mathrm{NT}$, and $\sigma_\mathrm{T}$, are the non-thermal, and the thermal velocity dispersions, respectively. $T_\mathrm{kin}$ is the kinetic temperature of the gas, which we assume $T_\mathrm{kin} = T_\mathrm{dust}$, $k_\mathrm{B}$ is the Boltzmann constant, $\bar{m}$ and $m_\mathrm{obs}$ refer to the mean molecular mass (2.37 a.m.u) and observed molecular mass (29 a.m.u for \ntwoh), respectively. We examine this non-thermal contribution with respect to the sound speed of the gas. This is referred to as the sonic Mach number,\footnote{Note that here we have determined the sonic Mach number ($\mathcal M_\mathrm{s}\,=\,\sigma _\mathrm{NT} / \sigma _\mathrm{T}$) using the 1 dimensional velocity dispersion ($\sigma_\mathrm{NT}$). However, this value can be converted to the 3 dimensional sonic Mach number by accounting for a factor of $3^{0.5}$; $\mathcal M_\mathrm{s,3D} = 3^{0.5} \mathcal M_\mathrm{s}$ (e.g. \citealp{palau15}).} or $\mathcal M_\mathrm{s}\,=\,\sigma _\mathrm{NT} / \sigma _\mathrm{T}$. We find $\mathcal M_\mathrm{s} = 1.7_{-0.8}^{+0.9}$ across the cloud sample (see Table\,\ref{tab:core_physprops}). \note{The uncertainty from the fitting procedure on $\sigma_\mathrm{obs}$ is typically $\sim$\,5\,per cent, which is taken as the uncertainty on $\sigma_\mathrm{v}$. This $\sim$\,5\,per cent uncertainty also applies to $\sigma _\mathrm{NT}$ and $\mathcal M_\mathrm{s}$ for the quiescent cores. For the star-forming sources we propagate the higher temperature uncertainty, and determine a $\sim$\,30\,per cent on $\sigma_\mathrm{NT}$, and $\sim$\,50\,per cent on $\mathcal M_\mathrm{s}$.}

The non-thermal motions could be indicative of turbulent motions within the cores, or the ordered global collapse of the cores (e.g. \citealp{kauffmann_2013}). Figure\,\ref{fig:hist_mach} shows histograms of $\mathcal M_\mathrm{s}$ for the whole sample, and for the star forming cores and quiescent (non-star forming) cores only. We find that the distribution for the star forming cores peaks towards higher values of $\mathcal M_\mathrm{s}$ compared to the quiescent (non-star forming) cores. Indeed, we determine median values of $\mathcal M_\mathrm{s}\,=\,1.57$ for the quiescent cores, and $\mathcal M_\mathrm{s}\,=\,$2.04 for star forming cores. We conduct a Kolmogorov-Smirnov test to determine the significance of this difference. We find a p-value of 0.11 for the samples, which can not reject the null hypothesis that the two samples are the same at a $<$\,10\,per cent accuracy. Thus, we cannot statistically confirm if the $\mathcal M_\mathrm{s}$ values from star-forming and quiescent cores are from the same distribution. Nonetheless, it is worth highlighting that a difference in the sonic Mach number for star forming and non-star forming cores across a similar size scale has already been noted within the literature, and attributed to feedback effects from embedded young-stellar objects (e.g. \citealp{sanchezmonge13}).  

\subsection{Stability assessment}\label{sec:stability}

In this section, we assess the stability of the cores against gravitational collapse. We consider that self-gravity is the only force causing the collapse of the cores (e.g. neglected any added pressure caused by the host cloud), and determine the balance against thermal support, thermal and turbulence support, and then consider any necessary magnetic fields required for further support against collapse.

\subsubsection{Thermal support}\label{sec:stability_therm}

We, firstly, investigate the support from gravitational collapse by the thermal pressure only. To do so, we determine the so-called Jeans mass, $M_\mathrm{J}$, which gives the maximum mass that can be supported by thermal pressure, and the Jeans length $\lambda_\mathrm{J}$, which gives the length scale of the fragmentation. The Jeans mass can be given as \citep{jeans_1902}, 
\begin{equation}
M_\mathrm{J} = \frac{\pi^{5/2} c_\mathrm{s}^3}{6 G^{3/2} \rho^{1/2}}, 
\end{equation}
where $\rho$ is the volume density of the core, and the sound speed is $c_\mathrm{s} = k_\mathrm{B}T_\mathrm{kin} / m_\mathrm{H} \mu_\mathrm{H_2} \sim 0.25$\,\kms\ at the median core dust temperature of 18\,K, and $G$ is the gravitational constant. We find values of the Jeans mass across the core sample of $M_\mathrm{J} = 0.8_{-0.1}^{+0.3}$\sol, and when using the background subtracted mass estimates: $M_\mathrm{J}^\mathrm{b} = 1.4_{-0.5}^{+0.9}$\,\sol. We compare these masses to the measured core masses, and find median ratios of $M/M_\mathrm{J} = 19.9_{-15.7}^{+45.7}$ ($M^\mathrm{b} /M_\mathrm{J}^\mathrm{b} = 2.7_{-2.3}^{+19.4}$). These ratios of $M/M_\mathrm{J}>1$ then show that the cores are potentially unstable to gravitational collapse if not additionally supported. These gravitationally unstable cores are potentially susceptible to further fragmentation. We estimate the corresponding Jeans length using, 
\begin{equation}
\lambda_\mathrm{J} = c_\mathrm{s} \left (  \frac{\pi}{G \rho}  \right ) ^{1/2}.
\label{eq:jeanslength}
\end{equation}
We find a value of the Jeans length across the sample of $\lambda_\mathrm{J} = 0.03_{-0.01}^{+0.01}$\,pc (or $6512_{-1756}^{+2439}$\,AU), and $\lambda_\mathrm{J}^\mathrm{b} = 0.06_{-0.02}^{+0.03}$\,pc (or $12022_{-3804}^{+6783}$\,AU). Comparing these values to the projected radius of the cores we find a ratio of $R_\mathrm{eff}/\lambda_\mathrm{J} = 1.36_{-0.54}^{+0.66}$, highlighting that these Jeans unstable cores could then fragment on size scales similar to the current observed core size scales. We will discuss this again later in this work, when we investigate the hierarchical structure of these cores using higher resolution datasets. 

\subsubsection{Thermal and turbulence support}\label{sec:stability_thermturb}

We now assess the balance of the total kinetic energy, $E_{\rm kin}$, including both the thermal and turbulent pressure against the gravitational potential energy, $E_{\rm pot}$. These energy terms can be equated to produce the commonly used virial parameter, $\alpha_{\rm vir}$ (e.g. \citealp{bertoldi_1992}). In the idealised case of a spherical core of uniform density supported by only kinetic energy (i.e. no magnetic fields), the virial parameter takes the form, 
\begin{equation}
\alpha_{\rm vir} = a \frac{5 \sigma_\mathrm{v}^{2} R_\mathrm{eff}}{G M},
\label{equ:virial}
\end{equation}
where $R_\mathrm{eff}$ is the effective radius of the core, $M$ is the total mass of the core, $\sigma_\mathrm{v}$ is the line-of-sight velocity dispersion; i.e. including both the thermal and turbulent broadening ($\alpha_{\rm vir}$ does not account for any systematic infall/outflow motions; see e.g. \citealp{kauffmann_2013}). The factor $a$ accounts for systems with non-homogeneous and non-spherical density distributions, and for a wide range of core shapes and density gradients takes a value of $a = 2\,\pm\,1$ (see \citealp{bertoldi_1992}). In the above, a value of $\alpha_{\rm vir} < 2$ indicates the core is sub-virial and may collapse, whereas for a value of $\alpha_{\rm vir} > 2$ the core is super-virial and may expand.

We find virial parameters across the sample of $\alpha_\mathrm{vir} = 0.7_{-0.4}^{+0.5}$ ($\alpha^\mathrm{b}_\mathrm{vir} = 2.3_{-1.5}^{+3.8}$). \note{The propagated error on $\alpha_\mathrm{vir}$ is $\sim$\,50 and $\sim$\,60\,per cent for the quiescent and star-forming cores, respectively.} We then find that the majority of cores within our sample have $\alpha_{\rm vir}$ less than 2 and, therefore, would appear to be bound and unstable to collapse. That said, these virial parameters are closer to a pressure balance than when only considering thermal pressure, which highlights the relative importance of the non-thermal motions in support against gravitational collapse. It should also be noted that this result appears sensitive to the background subtraction. 

We also assess the fragmentation of the cores using the total Jeans mass, $M_\mathrm{J,tot}$, which accounts for both the contribution of the thermal and non-thermal velocity dispersion. This can be calculated by substituting the $\sigma_\mathrm{v}$ for $c_\mathrm{s}$ in equation\,\ref{equ:virial} (e.g. \citealp{palau15, sadaghiani20}). We find values of $M_\mathrm{J,tot} = 4.4_{-3.2}^{+9.4}$\,\sol\ ($M_\mathrm{J,tot}^\mathrm{b} = 8.7_{-6.4}^{+14.6}$\,\sol), or as a ratio to the measured mass $M/M_\mathrm{J,tot} = 3.1_{-1.8}^{+9.4}$ ($M/M_\mathrm{J,tot}^\mathrm{b} = 0.5_{-0.4}^{+1.8}$). These values of the total Jeans mass are typically factors of a few higher than when accounting for only the thermal support (see section\,\ref{sec:stability_therm}), and more comparable to the measured masses. However, as shown by the virial parameter, values of $M/M_\mathrm{J,tot}>1$ highlight that the cores are likely to collapse and/or fragment unless further supported.

The parameters determined in this section for each source are given in Table\,\ref{tab:core_physprops}. We also note, we have investigated the virial state and total Jeans mass of the cores when using the \ntwoh\ cubes that have been corrected for the zero-spacing, i.e. feathered with the single-dish observations. When doing so we still find that the majority of the cores are kinematically sub-virial and have masses larger than their associated Jeans masses, suggesting that they may be susceptible to collapse and fragmentation if other means of support, e.g. magnetic fields, are not significant.

\subsubsection{Thermal, turbulence and magnetic support}\label{sec:stability_thermturbmag}
 
Lastly, we assess the relative importance of the magnetic field in support against gravitational collapse. To do so, we follow \citet{henshaw_2016} and calculate the virial parameter that includes the magnetic field contribution \citep{pillai_2011}, 
\begin{equation}
\alpha_{\rm B, vir} = a \frac{5 R_\mathrm{eff}}{G M} \left ( \sigma_\mathrm{v}^{2} - \frac{ \sigma_\mathrm{A}^{2} }{6} \right ),
\label{equ:virialmag}
\end{equation}
where the Alfv\'en velocity is $\sigma_\mathrm{A} = B (\mu_0 \rho)^{-1/2}$, in which $B$ is the magnetic field strength and $\mu_0$ is the permeability of free space. Here then we ask: how much magnetic field pressure is required in addition to turbulence and thermal pressure to support the cores against gravity? To do so, we set $\alpha_{\rm B, vir}=a=2$, and solve equation\,\ref{equ:virialmag} for $B$ for all cases where $\alpha_{\rm vir}<2$ (section\,\ref{sec:stability_thermturb}). We find values of $\sigma_\mathrm{A} = 0.91_{-0.45}^{+0.49}$\,\kms\ ($\sigma_\mathrm{A}^\mathrm{b} = 0.65_{-0.37}^{+0.36}$\,\kms), which correspond to Alfv\'en Mach numbers of $\mathcal M_\mathrm{A}\,=\,\sigma_\mathrm{A} / \sigma _\mathrm{T}= 3.76_{-1.88}^{+2.08}$ ($\mathcal M_\mathrm{A}^\mathrm{b} = 2.75_{-1.55}^{+1.48}$). The range of magnetic field strengths required for the stability of the $\alpha_{\rm vir}<2$ cores is then $B = 520_{-242}^{+470}$\,$\mu$G (or $B^\mathrm{b} = 271_{-181}^{+300}$\,$\mu$G).

We compare these measurements to the \citet{crutcher_2010} relation linking the magnetic field strength (determined from Zeeman splitting) and volume density, 
\begin{equation}
B_\mathrm{med} \approx \frac{1}{2} B_0 \left( \frac{n(\mathrm{H})}{n_0} \right) ^{2/3} \approx \frac{1}{2}  B_0 \left( \frac{2n_\mathrm{H_2}}{n_0} \right) ^{2/3},
\label{equ:Bmed}
\end{equation}
where $B_\mathrm{med} = B_\mathrm{max}/2$, and for $n(\mathrm{H})$\,$>$\,$n_0$ where $n_0 = 300$\,\cmcb, $n(\mathrm{H})=2n_\mathrm{H_2}$, and $B_0 = 10$\,$\mu$G. We find values of $B_\mathrm{med} = 1206_{-411}^{+572}$\,$\mu$G ($B_\mathrm{med}^\mathrm{b} = 543_{-238}^{+343}$\,$\mu$G), or $B/B_\mathrm{med}=0.49_{-0.23}^{+0.26}$ ($B^\mathrm{b}/B_\mathrm{med}^\mathrm{b}=0.39_{-0.20}^{+0.17}$). This shows that the magnetic field required for the additional support against gravitational collapse could then be more than achieved if these cores follow the \citet{crutcher_2010} relation, which is broadly consistent with the typical magnetic field strengths observed within molecular clouds \citep{pillai_2015, pillai_2016, soam_2019, tang_2019}. 

In a comparable resolution 3\,mm dust continuum study of the cores within the north portion of cloud H, \citet{henshaw_2016c} found that values of $B \sim 590$\,$\mu$G are required to stabilise the cores (or $B \sim 830$\,$\mu$G when accounting for the $\alpha_{\rm B, vir}=2$ imposed in this work). This value is broadly consistent with the range given by the standard deviation around the median value for the core sample (287 - 1000\,$\mu$G). In the southern portion of Cloud H studied in this work, we estimate a range of $B \sim 278 - 704$\,$\mu$G, suggesting that the magnetic field required for support across this whole filament is reasonably constant. Moreover, in their study of cores identified from N$_2$D$^+$\,(3$-$2) emission, \citet{tan_2013} and \citet{kong_2017} also estimated that $\sim$\,1\,mG B-fields were required for support against collapse. It is worth noting, however, that these magnetic  field strengths required for support against collapse are around an order of magnitude larger than the mean plane-of-the-sky magnetic field strength measured over parsec scales for Cloud H ($\sim$50\,$\mu$G; \citealp{liuT18}).

\section{Multi-scale analysis of cloud structure}\label{sec:analysis}

\subsection{Homogenised sample of literature cores}\label{sec:litomogenised}

We aim to make a comparison between the properties of the cores determined in this section to those presented within the literature (e.g. \citealp{kauffmann_2013, peretto_2013, sanchezmonge13}). Moreover, as the IRDC sample studied here has been the subject of several mm-band studies at varying angular resolutions (e.g. \citealp{rathborne_2006, liu_2018}), we also aim to study how the cores fragment over several orders of magnitude in size-scale.

The 10 clouds studied in this work have also been observed, at least in part, by the following studies: \citet[][henceforth R06]{rathborne_2006}, \citet[][henceforth H16]{henshaw_2016c}, \citet[][henceforth H17]{henshaw_2016d} and \citet[][henceforth L18]{liu_2018}. The frequencies, angular resolutions and source samples of each of these observations are given in Table\,\ref{tab:lit_obsprops}. To follow the fragmentation of the cores, we first standardise the parameters used to determine the physical properties of the literature sample. In doing so, we attempt to remove any systematic variations of the properties produced by the differing underlying assumptions imposed by each work. We create a single table containing the observed flux densities and effective angular radii (i.e. $R_\mathrm{eff} = \sqrt{A/\pi}$) from each of the literature samples (see Table\,\ref{tab:core_litprops}). We then recalculate the physical radius of each core assuming the cloud distances given in Table\,\ref{tab:cloudprops}, and masses using equation\,\ref{equ:mass} and the assumptions given in section\,\ref{sec:physprops}. We recalculate $\kappa_\nu$ for each set of observations using the frequencies given in Table\,\ref{tab:lit_obsprops}, and $\beta = 1.75$ \citep{battersby_2011} and $\kappa_0$\,=\,0.9\,cm$^{2}$\,g$^{-1}$ \citep{ossenkopf_1994}. We assume a constant dust temperature for all literature calculations of $T_\mathrm{dust}$=18\,K, which approximately corresponds to the mean dust temperature measured across the cores identified in this work (see section\,\ref{sec:physprops}). The standardised properties for the literature cores are provided in Table\,\ref{tab:core_litprops}.

It should be noted, the homogenisation of the literature dataset neglects variations in physical properties, such as dust-opacity, gas-to-dust ratio, and temperature, which could vary as a function of the size-scale. However, these properties have not been accurately measured for the IRDC sample across all scales. Hence, the homogenised comparison presented here is favoured for its simplicity over, e.g. arbitrarily varying the temperature or opacity as a function of size. Assuming dust temperatures of $T_\mathrm{dust}$=10\,K and 30\,K rather than 18\,K would cause the mass of a core identified at 3\,mm to vary by factors of 2.0 and 0.57, respectively (see section\,\ref{sec:physprops} for discussion of uncertainties on the temperature estimates of the star-forming sources). Additionally, assuming $\beta$ values of 1.5 and 2, rather than 1.75, would cause the mass of a core identified at 3\,mm to vary by factors of 0.79 and 1.25, respectively.

We define a referencing nomenclature for the structures within this catalogue based on the approximate scales over which the literature observations cover. For simplicity, we refer to the largest (0.1-1\,pc) structures from \citetalias{rathborne_2006} as clumps, the intermediate size (0.01-0.1\,pc) structures identified in this work and \citetalias{henshaw_2016c} as cores, and the smallest structures identified by \citetalias{liu_2018} and \citetalias{henshaw_2016d} as core-fragments. It should, however, be kept in mind that this is a simplification for referencing purposes, and is solely based on their approximate size scales with no bearing on their stellar population formation potential (e.g. \citealp{williams_2000}). Moreover, due to varying distances and physical properties across the cloud sample, there is some cross-over between these groups; e.g. for the closest clouds, several \citetalias{rathborne_2006} clumps have radii $<$0.1\,pc, and, therefore, could be classified as cores based on their size-scales.

\begin{table}
\centering
\caption{Observational properties of the literature sample (section\,\ref{sec:litomogenised}). Tabulated is the type of structure, reference, the telescope used to study the core samples, the clouds covered, and the wavelength and angular resolution of the observations (mean of major and minor axis of beam size). The abbreviated reference are \citet[][R06]{rathborne_2006}, \citet[][H16]{henshaw_2016c}, \citet[][H17]{henshaw_2016d} and \citet[][L18]{liu_2018}.}
\label{tab:lit_obsprops}
\begin{tabular}{cccccc}
\hline \hline
Structure & Ref. & Telescope & Sample & $\lambda$ & $\theta$ \\
 & & & & mm & $\mathrm{{}^{\prime\prime}}$ \\
\hline
Clump & \citetalias{rathborne_2006} & IRAM-30m & All & 1.3 & 11 \\
Core & \citetalias{henshaw_2016c} & IRAM-PdBI & H & 3 & 3.75\\
Core-frag. & \citetalias{henshaw_2016d} & ALMA & H(6) & 1 & 1\\
Core-frag. &\citetalias{liu_2018} & ALMA & Not  I/J & 1.3 & 1\\
Core & This work & ALMA & All & 3\ & 2.9 \\
\hline
\end{tabular}
\end{table}

\subsection{Mass- and linewidth-size relations}\label{sec:size}

\begin{figure}
\centering
    \includegraphics[width=\columnwidth]{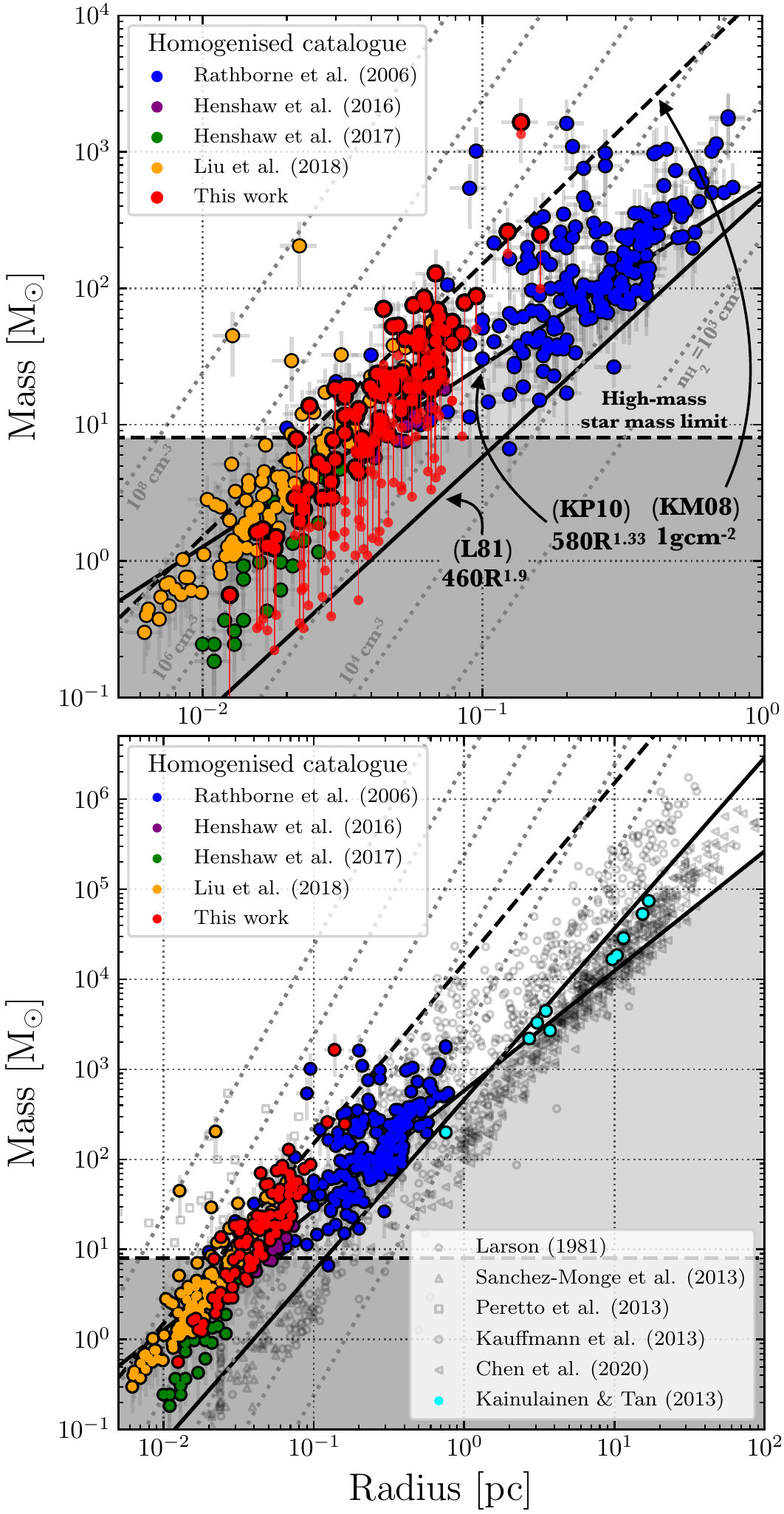}
    \vspace{-6mm}
    	\caption{A comparison of the mass and sizes for the cores determined in this work, to samples taken from the literature, and mass-size relations and thresholds for massive star formation. Shown as black-outlined red points are the cores studied in this work ($M$), which are connected to their corresponding lower background-subtracted masses ($M^\mathrm{b}$; see Table\,\ref{tab:core_physprops}). Highlighted as coloured points are the mass and sizes of the cores and/or clumps determined by \citet{liu_2018}, \citet{henshaw_2016c}, \citet{henshaw_2016d}, and \citet{rathborne_2006}, which have crossover with the IRDC sample examined in this work. The properties of these cores have been recalculated using the same assumptions outlined in this work (section\,\ref{sec:litomogenised}). (lower panel) Shown as grey open points are the properties taken directly from \citet{larson_1981}, \citet{kauffmann_2013}, \citet{peretto_2013}, \citet{sanchezmonge13}, and \citet{chen_2020}. Overlaid as diagonal solid black lines are the mass-radius relations taken from \citet[][L81]{larson_1981}, and high-mass star formation thresholds taken from \citet[][KM08]{krumholz_2008} and \citet[][KP10]{kauffmann_2010}. The \citetalias{kauffmann_2010} relation has been scaled by a factor of 1.5 to match the dust opacity used throughout this paper. The horizontal dotted black line shows the mass threshold of high-mass stars (>8\,\sol). The dashed diagonal grey lines show constant number densities (as labelled).} 
    	\label{fig:mr}
\end{figure}

A simple first analysis using our extensive core catalogue is to plot the masses, velocity dispersion and virial states as a function of size (e.g. \citealp{larson_1981}).

\subsubsection{Mass-size relation}

Figure\,\ref{fig:mr} presents the mass as a function of the effective radius for the core sample identified within this work compared to sources taken from the literature. Shown as black-outlined red points are the masses ($M$) and as red points are the background-subtracted masses ($M^\mathrm{b}$) determined in section\,\ref{sec:physprops} (see  Table\,\ref{tab:core_physprops}). Shown as the coloured points are the masses and effective radii from the homogenised literature core catalogue (section\,\ref{sec:litomogenised}). For a representative comparison, we also plot on Figure\,\ref{fig:lwr} the core samples from \citet[][K13]{kauffmann_2013}, \citet[][P13]{peretto_2013}, and \citet[][SM13]{sanchezmonge13}. 

Figure\,\ref{fig:mr} shows that smaller cores are typically less massive, within an order of magnitude scatter in mass for any given effective radius. Shown as diagonal dotted grey lines are constant mean number densities. We see that the core fragments have $n_\mathrm{H_2}\sim10^{6-7}$\,cm$^{-3}$, whereas the clumps have $n_\mathrm{H_2}\sim10^{4-5}$\,cm$^{-3}$. Showing that the smallest cores are less massive, yet significantly denser than their more massive counterparts. The larger clumps could, however, also achieve larger volume densities at the same size scales of the smaller cores and core fragments, in the likely case that they are centrally concentrated. This could imply that the derived volume density of the larger clump is smaller because most of their volume is at lower densities. 

Overlaid on Figure\,\ref{fig:mr} are several commonly adopted thresholds for massive-star formation, which we can compare to the observed cores. The horizontal dashed black line, and dark shaded region is the limit above which a star can be considered high-mass ($\sim$\,8\sol). Note that this mass limit does not account for a star formation efficiency, and, therefore, represents a lower limit for the required mass within a structure. Ultimately, any core with a mass lower than this threshold can not in its current state form a high-mass star. Also overlaid on Figure\,\ref{fig:mr} as several black diagonal lines are often quoted mass-size relations. 

The first of these mass-size relations is from \citet[][henceforth L81]{larson_1981}, and is given as $M\,=\,460R^{1.9}$ where the normalisation approximately relates to the mass surface density of their sample of local sources ($\Sigma=460/\pi$\,\sol\,pc$^{-2}$), and $M$ and $R$ are the mass and radius in units of \sol\ and parsec.\footnote{\note{Determined by equating $\sigma_\mathrm{v}\,=\,1.1\,(L)^{0.38}\,=\,1.1\,(2R)^{0.38}$ and $\sigma_\mathrm{v}\,=\,0.42\,(M)^{0.2}$, where $L$ is the core diameter, and solving for $M$ and $R$ \citep{larson_1981}.}} \note{We see that almost all of the clumps and cores within the homogenised core catalogue (all coloured data points) sit above the \citetalias{larson_1981} relation. The reason for this difference, which can be up to an order of magnitude for the smallest cores, could be a result of several factors.} 
\note{Firstly, there are several systematics to consider. \citetalias{larson_1981} defined the size as the is maximum linear extent of the structure (i.e. larger than $r_\mathrm{eff}$ used here). Secondly, the \citetalias{larson_1981} mass estimates are determined from fundamentally different observations than used in this work. In this work, we analysis the dust continuum emission to estimate the masses, yet \citetalias{larson_1981} used molecular line (e.g. CO) excitation arguments and simple assumptions about the geometry (see e.g. \citealp{lada20} for further discussion of how different mass tracers can cause scatter in the mass-size relation). Thirdly, the ALMA continuum observations presented within this work suffer from spatial filtering, which may affect this comparison to sizes and masses determined primarily from single-dish observations.}

\note{With the above systematics in mind, it is then worth considering that there may be an alternative interpretation of this result. On the largest scale (>1\,pc), we see that the clouds are in broad agreement with the \citetalias{larson_1981} relation (see lower panel of Figure\,\ref{fig:mr}).
The increase in $M$ relative to \citetalias{larson_1981} seen at smaller $r_\mathrm{eff}$ (<0.1\,pc) could then be explained by the fragmented structure of the clouds, and, on the smallest scales, the density profiles of the cores themselves (see e.g. \citealp{kauffmann10, kauffmann10b, ballesterosparedes12} for discussion). The \citetalias{larson_1981} relation fails to account for much of this complex structure we now know exists within molecular clouds (see e.g. \citealp{mckee_2007} for references showing significant scatter in mass-size relations), and seen in our observations (see Figure\,\ref{fig:mom_maps}). It is then not all too surprising that we find this deviation, particularly as here investigate scales below that probed by \citetalias{larson_1981}.}

Also over-plotted on Figure\,\ref{fig:mr} are the high-mass star formation relations proposed by \citet[][henceforth KM08]{krumholz_2008} and \citet[][henceforth KP10]{kauffmann_2010}. When determining their relation, \citetalias{kauffmann_2010} reduced the \citet{ossenkopf_1994} dust opacity used for the mass determination by a factor of 1.5 (see \citealp{kauffmann10}). Following \citet{dunham11}, we determine the relation to account for the dust opacity used throughout this work (see section\,\ref{sec:physprops}). The \citetalias{kauffmann_2010} scale relation is hence given as $M\,=\,580\,R^{1.33}$, where the $M$ and $R$ are the mass and radius in units of \sol\ and parsec. \note{It is worth just briefly noting here again, that the comparison to any empirical mass-size threshold may be complicated by systematics, such as the filtering characteristics that differ between observations (e.g. missing extended flux) and mass determination methodologies. For example, here \citetalias{kauffmann_2010} used a dendrogram analysis, and based the relation on the non-background subtracted mass estimates, yet we know that mass estimates can still vary depending on the contribution from fore/background emission. Hence, we emphasise caution when drawing any firm conclusion for the high-mass star-forming potential purely from such empirical mass-size thresholds (also see discussion in section\,\ref{sec:frag_structure}).}

The analytically determined mass surface density threshold from \citetalias{krumholz_2008} is given as $\Sigma\,\sim\,1$g\,cm$^{-2}$, which is based on a model where fragmentation of massive cores is inhibited by radiative heating from surrounding lower-mass protostars. The value of the mass surface density is required such that the lower-mass protostars have high mass accretion rates, and hence are luminous enough to sufficiently heat the massive core. The \citetalias{krumholz_2008} threshold approximately corresponds to $M\,=\,15000\,R^{2}$, in units of \sol\ and parsec. 

We find that the core sample sits around these two threshold relations, with the clumps scattering around the \citetalias{kauffmann_2010} relation. The light grey shaded region on Figure\,\ref{fig:mr} represents the parameter space below the $\sim$\,8\sol\ threshold for high-mass star formation. We find that a several of cores within the size-scale of $\sim$\,0.05\,pc sit above this shaded region, and hence would appear to be able to form a high-mass star when assuming no mass-loss or further fragmentation. We return the investigation of these sources as potential high-mass star forming regions later in section\,\ref{sec:hmcores}.



\subsubsection{Linewidth-size relation}

\begin{figure}
\centering
    \includegraphics[width=\columnwidth]{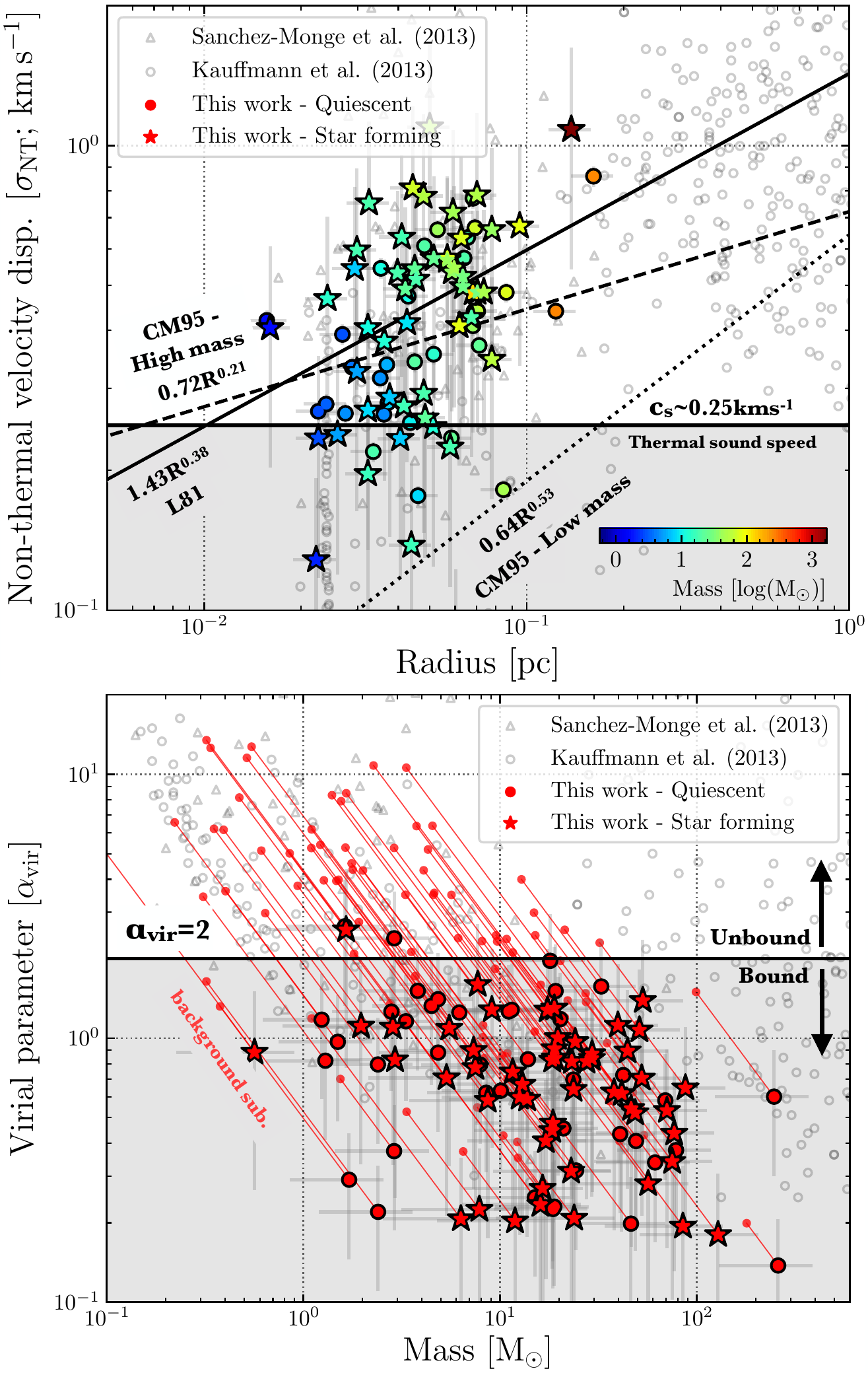}
    \vspace{-6mm}
    	\caption{Dynamical state of the identified cores. The upper panel shows the non-thermal velocity dispersion as a function of the effective radius, and the lower panel shows the virial parameter as a function of the mass. Shown as black-outlined circle and star markers are the cores studied in this work that have been identified as quiescent and star-forming, respectively (see section\,\ref{sec:coreident}). The points in the upper panel have been coloured according to their measured masses, as shown by the inset log-scale colourbar. In the lower panel, also shown by the red points without an outline are the corresponding background-subtracted properties (see Table\,\ref{tab:core_physprops}). Shown as grey open circles in both panels is the core samples from \citet[][K13]{kauffmann_2013} and \citet[][]{sanchezmonge13}. The diagonal and horizontal black lines shown in the upper panel represent the \citet[][L81]{larson_1981}, \citet[][CM95]{caselli_1995b} low- and high-mass linewidth-size relations, and the thermal sound speed of the gas ($c_\mathrm{s} = (k_\mathrm{B} T_\mathrm{kin}) / (m_\mathrm{H} \mu_\mathrm{p}) \approx 0.25$\,\kms\ for $\mu_\mathrm{p} = 2.37$ and $T_\mathrm{kin}=18$\,K). The horizontal black line shown in the lower panel represents the limit between ($\alpha_{\rm vir} < 2$) a bound core, and ($\alpha_{\rm vir} > 2$) unbound core with non-homogeneous and non-spherical density distribution \citepalias{kauffmann_2013}.} 
    	\label{fig:lwr}
\end{figure}

We now investigate how the dynamical state of the core sample depends on size scale and mass. In Figure\,\ref{fig:lwr}, we plot the non-thermal velocity dispersion (upper panel) as a function of effective radius. Here we only show the cores determined in this work (red points), as there is only a partial determination of the velocity dispersion for the homogenised literature sample. Moreover, where such measurements have been made for the homogenised literature sample, such a comparison is complicated by the use of different molecular lines, which may originate from fundamentally different gas properties and hence different regions within the cores. Bearing this in mind, on Figure\,\ref{fig:lwr} we also show the core samples from \citetalias{sanchezmonge13} and \citetalias{kauffmann_2013} (see their discussion of various molecular line probes used to compile this sample). We find that the cores identified in this work are in reasonable agreement within the scatter of those from the literature. Also overlaid as a diagonal solid black line on Figure\,\ref{fig:lwr} is the \citetalias{larson_1981} linewidth-size relation, which is given as $\sigma_\mathrm{v}\,=\,1.1\,L^{0.38}=\,1.1\,(2R)^{0.38}=\,1.43\,R^{0.38}$ where $L$ is the core diameter, and $\sigma_\mathrm{v}$ and $R$ are in units of \kms\ and parsec. \note{Note that the \citetalias{larson_1981} relates to the total velocity dispersion, as opposed to the non-thermal velocity dispersion plotted in Figure\,\ref{fig:lwr}. We expect that this may cause a minor systematic, as the contribution of the thermal velocity dispersion is only small for the majority of cores (see section\,\ref{sec:dynprops}).} Moreover, we show the linewidth-size relations determined for both low- and high-mass within Orion by \citet[][CM95]{caselli_1995b}. These are given by $\sigma_\mathrm{NT}\,=\,0.64\,R^{0.53}$ for low mass, and $\sigma_\mathrm{NT}\,=\,0.72\,R^{0.21}$ for high mass cores, again in units of \kms\ and parsec \note{(note the conversion to velocity dispersion from the line-width provided in \citetalias{caselli_1995b})}. We find that the cores have range across all of the plotted line-width size relations, yet mostly cluster around the \citetalias{larson_1981} and \citetalias{caselli_1995b} high-mass relations. We have highlighted in the upper panel of Figure\,\ref{fig:lwr} the cores that are quiescent and star-forming, and colour the points by their measured masses. We find no preference to any of the line-width size relations with the core masses or state of star formation. We find several cores have very narrow velocity dispersions, and sit significantly below the \citetalias{larson_1981} relation, close to the \citetalias{caselli_1995b} low-mass relation. Shown as the grey shaded region on Figure\,\ref{fig:lwr} is $\sigma_\mathrm{NT}<c_\mathrm{s}$, or the regime where dynamical motions within the core are sub-thermal. Several cores from the \citetalias{kauffmann_2013} are also seen within this regime.

The lower panel of Figure\,\ref{fig:lwr} shows the virial parameter as a function of mass for the cores identified in this work. Shown as red circles and red circles outlined in black are the masses and virial parameters determined with and without background subtraction, respectively. Highlighted as a black horizontal line is $\alpha_{\rm vir} = 2$, which represents the boundary between gravitational bound and unbound cores (for a given core structure). We find that the majority of cores identified within this work have $\alpha_{\rm vir} \ll 2$, and, therefore, are expected to collapse without additional support. We find that when accounting for the background subtraction, many cores do move into the unbound regime ($\alpha_{\rm vir}>2$). We find no clear trend between the virial parameters and the mass or star-forming state of the cores.

\subsection{Fragmentation}\label{sec:frag}

 \begin{figure*}
 \centering
     \includegraphics[width=\textwidth]{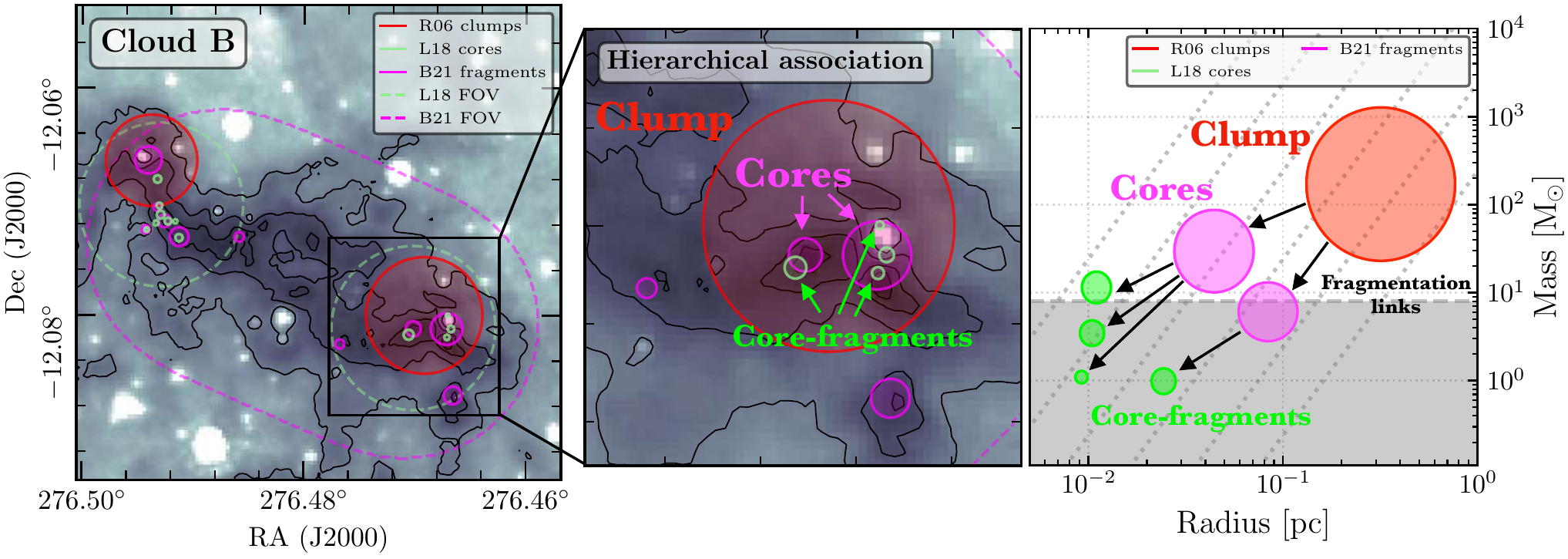}
     	\caption{A schematic diagram showing the fragmentation linking procedure, which we use to determine a hierarchy of structures similar to a dendrogram across multiple datasets covering our cloud sample (section\,\ref{sec:frag}). The left panel shows the near- and mid-infrared extinction derived mass surface density map for Cloud B in greyscale, overlaid with circles representing the position and sizes of the cores contained within the homogenised catalogue (section\,\ref{sec:litomogenised}). The red circles show the \citet[][R06]{rathborne_2006} clumps, the solid magenta circles show the cores identified in this work (B21), and the solid green circles show the core-fragments identified by \citet[][L18]{liu_2018}. The dashed magenta and green lines show the field of view (FOV) of the ALMA observations presented in this work and in \citetalias{liu_2018}, respectively. The centre panel shows a zoom-in of the MM2 \citetalias{rathborne_2006} core within Cloud B, which is highlighted on the left panel by a solid black box. Here the clump, core and core-fragment structures have been labelled. The right panel shows the mass as a function of radius for the example MM2 region. Here we show the fragmentation links between the host clump, and the cores and core fragments contained within its boundary.}
     	\label{fig:frag_toy}
 \end{figure*}

 \subsubsection{Determination of hierarchical structure}\label{sec:frag_structure}

 In this section, we aim to follow and connect the mass distribution through fragmentation of the complete homogenised clump-core-core fragment sample. In effect, what we create in this analysis is a hierarchy of structures similar to a dendrogram, yet we recover many magnitudes in spatial scale by combining various sets of observations towards the same sources taken at differing spatial resolutions. To create this structure for each of the IRDCs studied here, we first order the datasets in the homogenised structure catalogue by decreasing angular resolution (see Table\,\ref{tab:lit_obsprops}). We then take the largest scale clumps and the intermediate scale cores for a given IRDC (e.g. \citetalias{rathborne_2006} and this work), and determine if the positions of the cores are contained within the boundaries of the clumps. This step is then repeated decreasing in scale; e.g. with the cores from this work and core-fragments from \citetalias{liu_2018}. The result of this procedure is a structure catalogue with links between clumps that contain cores, and the cores that contain core-fragments. Figure\,\ref{fig:frag_toy} shows a schematic diagram of this fragmentation procedure within the MM2 clump within Cloud B, and how the determined fragmentation links can be interpreted through the mass-radius parameter space.

 We highlight that the fragmentation analysis shown in Figure\,\ref{fig:frag_toy} includes three clumps associated within mm-bright star-forming regions (Cloud C MM14, Cloud I MM3, and Cloud F MM9). These sources are particularly bright, and add confusion to the core identification in lower resolution observations from \citetalias{rathborne_2006}; such that e.g. close-by lower brightness cores merge with the bright source. Moreover, these embedded sources can produce a large increase in temperature, and so our masses may be overestimated. The resultant large uncertainty on the mass estimate can, for example, cause \citetalias{rathborne_2006} clumps to have masses that appear smaller than the contained cores. These have been highlighted on Figure\,\ref{fig:fragsep}, and their hierarchical structure should be taken with caution.

\begin{figure}
 \centering
     \includegraphics[width=\columnwidth]{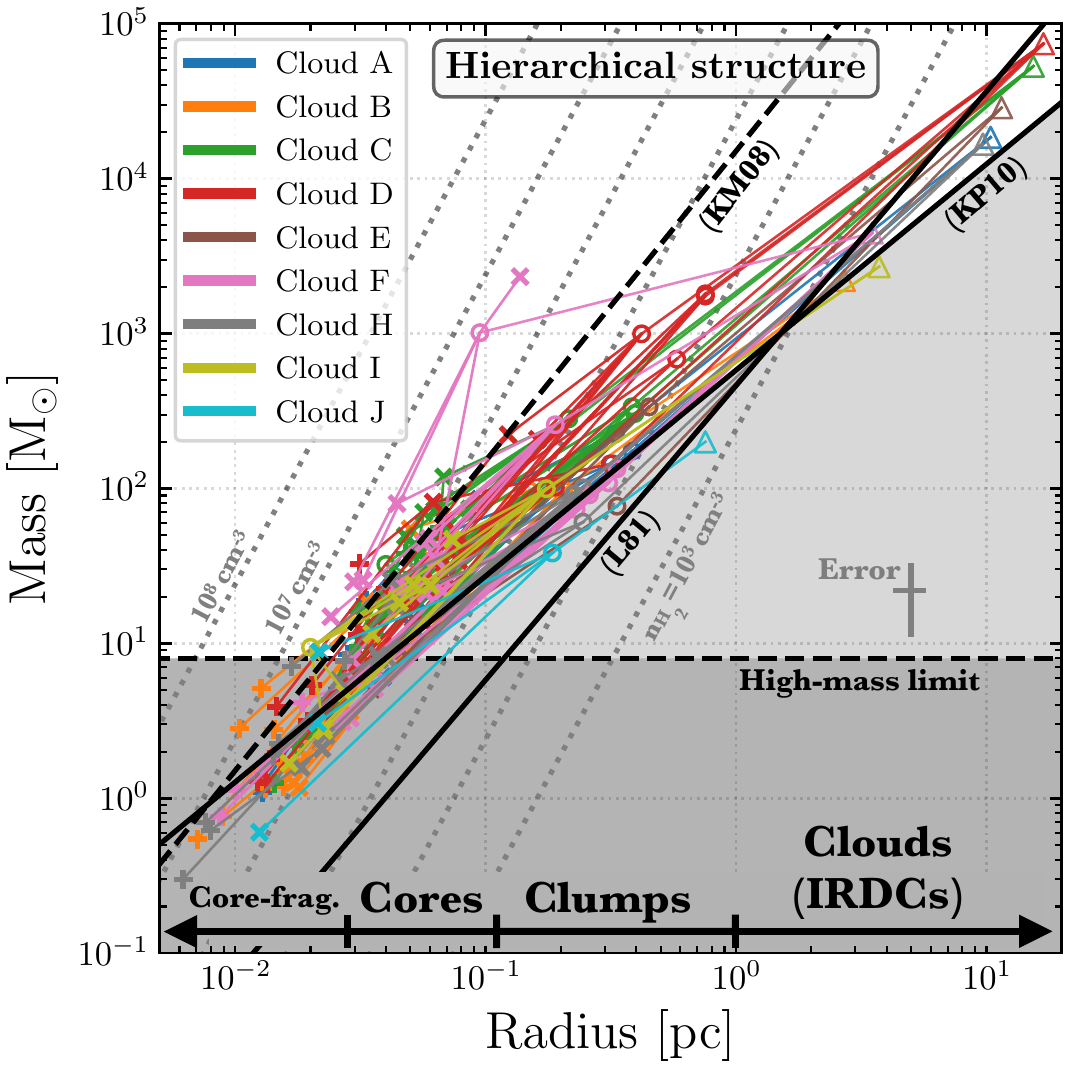}
     	\caption{The clump-to-core fragment masses within each cloud as a function of the size-scale. The circles, crosses and plus sign markers represent the clumps from \citetalias{rathborne_2006}, cores from this work and the clumps from \citetalias{liu_2018}. The straight lines connect the symbols for each core/core-fragment to the larger, host clump/core and then cloud (see Figure\,\ref{fig:fragsep} in the appendix for all clouds plotted separately). The lines and symbols have been coloured by cloud as indicated in the legend located in the upper left. The dotted diagonal grey lines show constant number densities (as labelled), and the black lines show the high-mass star formation thresholds shown in Figure\,\ref{fig:mr}. \note{In the lower right, we show a representative uncertainty range $\sim$\,15 and $\sim$\,50\,per cent on the radius and mass, respectively.}} 
     	\label{fig:frag}
 \end{figure}

 Figure\,\ref{fig:frag} shows the mass as a function of radius for all the homogenised catalogue. Here the circles, crosses and plus sign markers represent the clumps identified by \citetalias{rathborne_2006}, the cores from this work and the core-fragments from \citetalias{liu_2018}, respectively. We highlight the size-scales that correspond to the cores ($<0.1$\,pc) and core-fragments ($<0.01$\,pc) that are expected to form single stars, and the 0.1 to 1\,pc scale clumps that fragment into multiple cores that are expected to form several stars (e.g. \citealp{williams_2000}). These lines show the connections according to their fragmentation (see appendix for all clouds plotted separately). We find that all the cores identified within this work originate from the larger scale cores identified by \citetalias{rathborne_2006}. Moreover, in many cases, several cores originate from the same parent \citetalias{rathborne_2006} clump. We find that on smaller scales the cores identified here also fragment further, and in several cases also into multiple core-fragments. This further highlights the hierarchical nature of the interstellar medium (ISM), where structures can fragment at increasingly smaller scales. 

 Following the individual hierarchies on Figure\,\ref{fig:frag}, we see that the connected structures can cross the \citetalias{krumholz_2008} and \citetalias{kauffmann_2010} high-mass star formation thresholds at different scales. At large scales ($\sim$\,0.5\,pc), we see that clumps can have a moderate number density ($10^{4-5}$\,cm$^{-3}$), and sit around the \citetalias{kauffmann_2010} relation. At intermediate scales (0.1--0.5\,pc), the cores have a higher density ($10^{5-6}$\,cm$^{-3}$), and are comparable to both the \citetalias{krumholz_2008} and \citetalias{kauffmann_2010} relations. At the smallest scale ($\sim$\,0.01\,pc), the core-fragments lie either below the mass (<8\,\sol) threshold, or again under both the massive star formation relations. This then highlights the difficultly of these thresholds in predicting the high-mass star-forming potential of a region across any given spatial scale.

 \subsubsection{Identification of potential high-mass star-forming cores}\label{sec:hmcores}

 \begin{figure}
 \centering
     \includegraphics[width=\columnwidth]{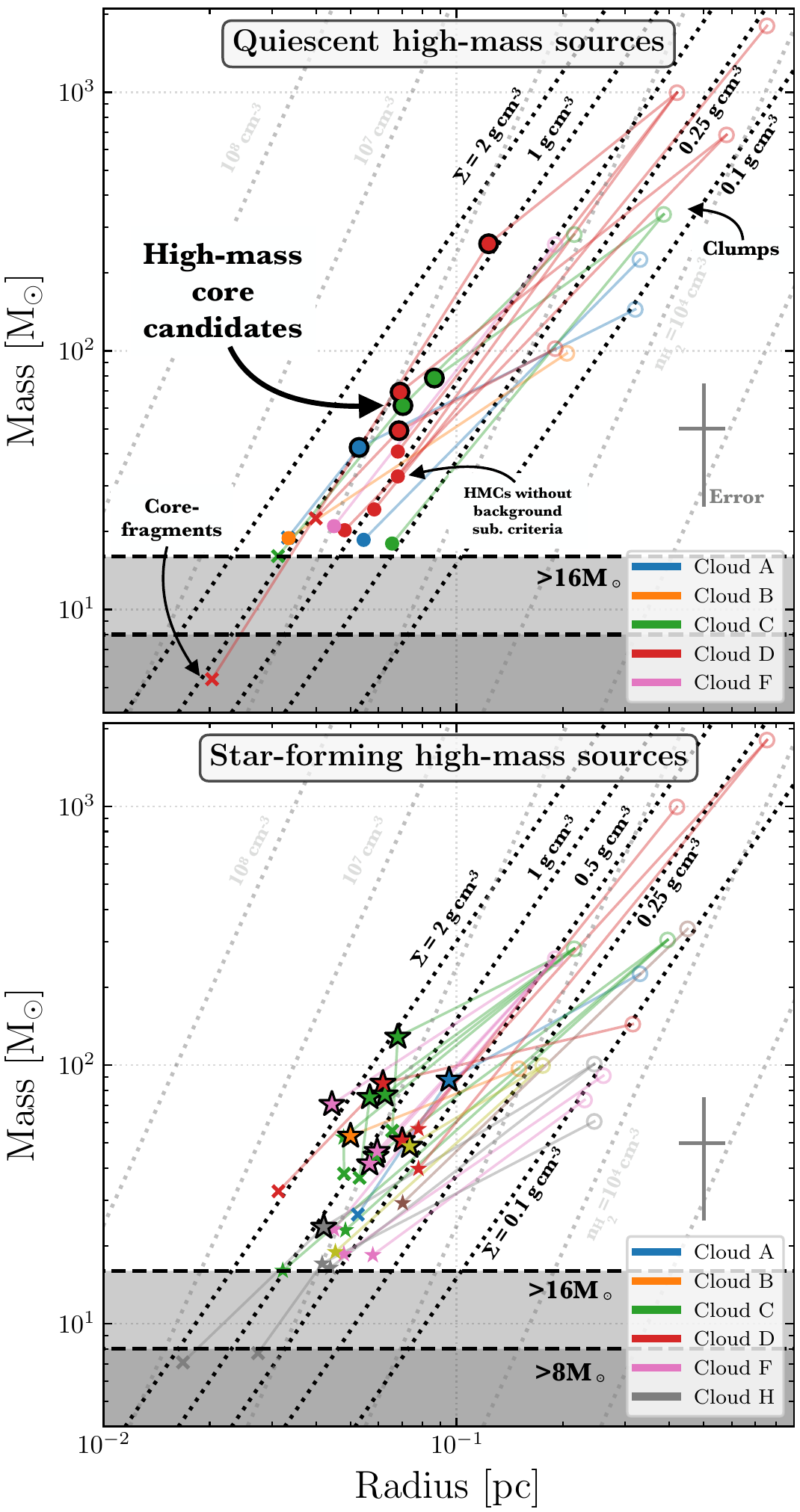}
     	\caption{The 19 high-mass quiescent and star-forming core candidates (HMCs; see section\,\ref{sec:hmcores}). As in Figure\,\ref{fig:frag}, we show the mass fragments within each cloud as a function of the size-scale. We highlight with black-outlined symbols the cores identified within this work, which have been singled out as high-mass ($M_\mathrm{c}>16$\,\sol\ and $M_\mathrm{c}^\mathrm{b}>16$\,\sol) without further fragmentation. We plot separately the cores that have been classified as quiescent and star-forming in the upper and lower panels, respectively. We also show as filled symbols with no outline cores identified in this paper that do not meet the background-subtracted mass threshold ($M_\mathrm{c}^\mathrm{b}>16$\,\sol). The \citet{rathborne_2006} clumps and \citet{liu_2018} core-fragment shown as faded open circles and crosses, respectively. The dotted diagonal faded grey lines show constant number densities (as labelled), and the black dotted diagonal lines show constant mass surface density ($\Sigma$; as labelled). \note{In the lower right, we show a representative uncertainty range $\sim$\,15 and $\sim$\,50\,per cent on the radius and mass, respectively.}} 
     	\label{fig:frag_hm}
 \end{figure}

 In this section, we attempt to use the hierarchy within each cloud to identify the cores that present the best candidates for the early stages of high-mass star formation. To do so, first, we impose that both the mass and the background-subtracted mass of the cores identified in this work, $M_\mathrm{c}$ and $M_\mathrm{c}^\mathrm{b}$, must be $>16$\,\sol, hence separating those cores that have enough mass to form at least a high-mass star when assuming a star formation efficiency of $\sim$\,50 per cent (e.g. see \citealp{tanaka17, liu20}). Secondly, we restrict the sample by the number of core-fragments, $n_\mathrm{cf}$, contained within each core. Cores with $n_\mathrm{cf}=0$ are included, as these represent the lowest level of the hierarchy; i.e. those that are high-mass, yet do not appear to fragment. Cores with $n_\mathrm{cf}=1$ are also included, as the core-fragment corresponds to a central emission peak, and hence these also represent cores that do not fragment. Cores with $n_\mathrm{cf}>1$ are included only if at least one of the core-fragments has a mass of $M_\mathrm{cf}>16$\,\sol. \note{We stress once more that the cores could have a complex three-dimensional, density and temperature structures that could influence this analysis. For example, the emission peak(s) could correspond to a central density peak, or, in the case of a star-forming sources, an increase in temperature (e.g. those with a 70\micron\ point source; see section\,\ref{sec:physprops}). A more in-depth study of the kinematic and chemical composition of these cores will be used to address this in the future, and here we make the simple, general assumption that the identified hierarchical structure is due to the physical fragmentation.} We separate this high-mass core sample by those with and without signs of ongoing star formation (i.e. have no associated infrared point source emission; see section\,\ref{sec:physprops}). 

 Out of the sample of 96 cores used for this analysis, we find that 19 cores satisfy the above requirements as potential candidates for high-mass star formation. Out of these, we find that 6 show no signs of active star formation, whilst 13 have signs of active star formation.\footnote{The quiescent high-mass cores are: A3c3, C2c1, C1c1, D8c1, D6c5, D6c4. The high-mass star-forming cores are: A1c1/2, B2c10, C2c2, C2c3/5, C2c6, D5c7, C6c1/2, D9c1/2, F4c5, F4c8, F4c10, H3c3, I1c1.} Figure\,\ref{fig:frag_hm} shows the mass and radius of the high-mass cores with the connection to their host clumps and contained core-fragments. We plot separately the cores that have been classified as quiescent and star-forming in the upper and lower panels, respectively. For reference, we also show the cores identified in this analysis that do not meet the background-subtracted mass threshold of ($M_\mathrm{c}^\mathrm{b}>16$\,\sol).

 We first focus on the 6 quiescent high-mass cores. We see that these are distributed across clouds A, C, and D, which contain 1, 2, and 3 core(s), respectively. The positions of these quiescent cores are shown on Figure\,\ref{fig:mom_maps} with cyan contours. We find that these have radii ranging from 0.05\,pc to 0.12\,pc, and masses from 42\sol\ to 260\sol\ (background subtracted masses of 21 to 180), which correspond to mean mass surface densities of 0.7\,g\,cm$^{-2}$ to 1.1\,g\,cm$^{-2}$. We have determined the dynamical properties of these cores. We find sonic Mach numbers across this sub-sample from 1.9 to 2.9 (median 2.0), and virial parameters from 0.14 to 0.73 (median 0.39). This then highlights that these cores contain trans-sonic non-thermal motions, are predominately kinematically sub-virial and require moderate magnetic field strengths of 780\,$\mu$G to 1380\,$\mu$G (median 976\,$\mu$G) for support against collapse. Without magnetic support, these cores would then be expected to form high-mass stars on the scale of a free-fall time, which we calculate ranges between 30,000\,yr and 50,000\,yr for the sample.

 The 13 high-mass cores that show signs of on-going star formation have radii ranging from 0.04\,pc to 0.10\,pc, and masses in the range of 24\sol\ to 129\sol that correspond to mass surface densities of 0.6\,g\,cm$^{-2}$ to 2.4\,g\,cm$^{-2}$. It should be kept in mind that the presence of star formation within these cores most likely means that the assumed dust temperature measurement is too low, and because of this these mass estimate have a larger associated uncertainty of $\sim$\,30\,per cent (section\,\ref{sec:physprops}). We find Mach numbers for these sources of 1.7 to 4.5 (median 2.4), virial parameters from 0.18 to 1.39 (median 0.54), which would require magnetic field strengths of 592\,$\mu$G to 2570\,$\mu$G (median 934\,$\mu$G) for additional support. The properties of this sub-sample are then generally higher than the quiescent high-mass core sample.

 We now assess how the properties of the high-mass cores compare to the modes of high-mass star formation introduced in section\,\ref{sec:intro}. The turbulent core accretion theory makes predictions for the sizes of massive pre-stellar cores (and thus early-stage cores) in IRDC environments. \citet[][their equation 20]{mckee_2003} propose that the radius of the core ($R_\mathrm{c}$) is linked to the core mass ($M_\mathrm{c}$) and the mean mass surface density of the cloud ($\Sigma_\mathrm{cl}$): $R_\mathrm{c} = 0.057 (R_\mathrm{c} / 60\mathrm{M_\odot})^{1/2} \Sigma_\mathrm{cl}^{-1/2}$\,pc. This then predicts that for a typical mass surface density across the cloud sample of $\Sigma_\mathrm{cl}\sim$\,0.1\,g\,cm$^{-2}$ \citep{kainulainen_2013}, a 16\sol\ core would have a radius of $\sim$0.09\,pc, respectively. Using the upper limit of the mass surface density within the clumps instead of the average across the cloud, we predict a 16\sol\ core would have a radius of 0.04\,pc. These values are broadly comparable to the measured size range of high-mass core sample, therefore the existence of these structures is consistent with above predictions from the core accretion theory.

Several mechanisms have been proposed to support high-mass cores against significant fragmentation in core accretion theory. The aforementioned \citet{krumholz_2008} relation predicts that the suppression of fragmentation is a result of the warming from a population of lower mass protostars. This requires the environments surrounding the young high-mass stars to have high mass surface densities ($\Sigma\sim$1\,g\,cm$^{-2}$), such that the protostars have high accretion rates. It is then interesting to consider that the cores that show signs of active star formation typically (6 out of 13) have mass surface densities of $>$1\,g\,cm$^{-2}$. An alternative mechanism for the suppression of fragmentation is from strong magnetic field strengths, which we estimate should be of the $\sim$\,1mG to support against gravitational collapse. Magnetic field strengths similar of this order have been previously observed within molecular clouds \citep{pillai_2015, pillai_2016, soam_2019, tang_2019, liuJ20}.

It should be noted that there are high-mass cores that have been identified in this work that do fragment on smaller scales. We find that 3 cores meet the above mass requirements, yet fragment into more than two lower mass core-fragments.\footnote{Cores B1c1/2/3, D5c5/6, F1c1} These would be interesting candidates to follow-up in the context of lower-mass star formation, or future high-mass cores.


Lastly, we highlight several future observations that would help assess the high-mass star-forming potential of these cores. Firstly, cores C1c1 (quiescent) and I1c1 (star-forming) were not covered by the \citet{liu_2018} observations, and, therefore, it would be interesting to investigate if these fragment in high spatial resolution observations. Second, it would be interesting to investigate if any of these sources contain signs of lower mass star formation that would be evident within the infrared emission, such as searching for outflows. A comparison to the positions outflows in Cloud C shows that the high-mass (quiescent) core C1c1 could indeed already contain embedded lower mass protostars \citep{feng16a,feng16b,kong19b}. Third, the quiescent high-mass core C2c1 has been observed at high-resolution by both \citet{zhang_2015} and \citet{kong_2017}. These studies find that this core contains $\sim$\,30\sol\ on the scale of $\sim$\,10$^{-3}$\,pc, has no shock or outflow tracer emission and has significant N$_2$D$^+$\,(3-2) emission. Indicating that core C2c1 is chemically young, and a particularly interesting target for follow-up studies of the earliest stages of high-mass star formation. Lastly, measurements of the magnetic field strength across the cores would be useful in assessing their stability.

 \subsubsection{From clouds ($\sim$\,1\,pc scales) to cores ($\sim$\,0.1\,pc scales)}

 \begin{figure*}
 \centering
     \includegraphics[width=\textwidth]{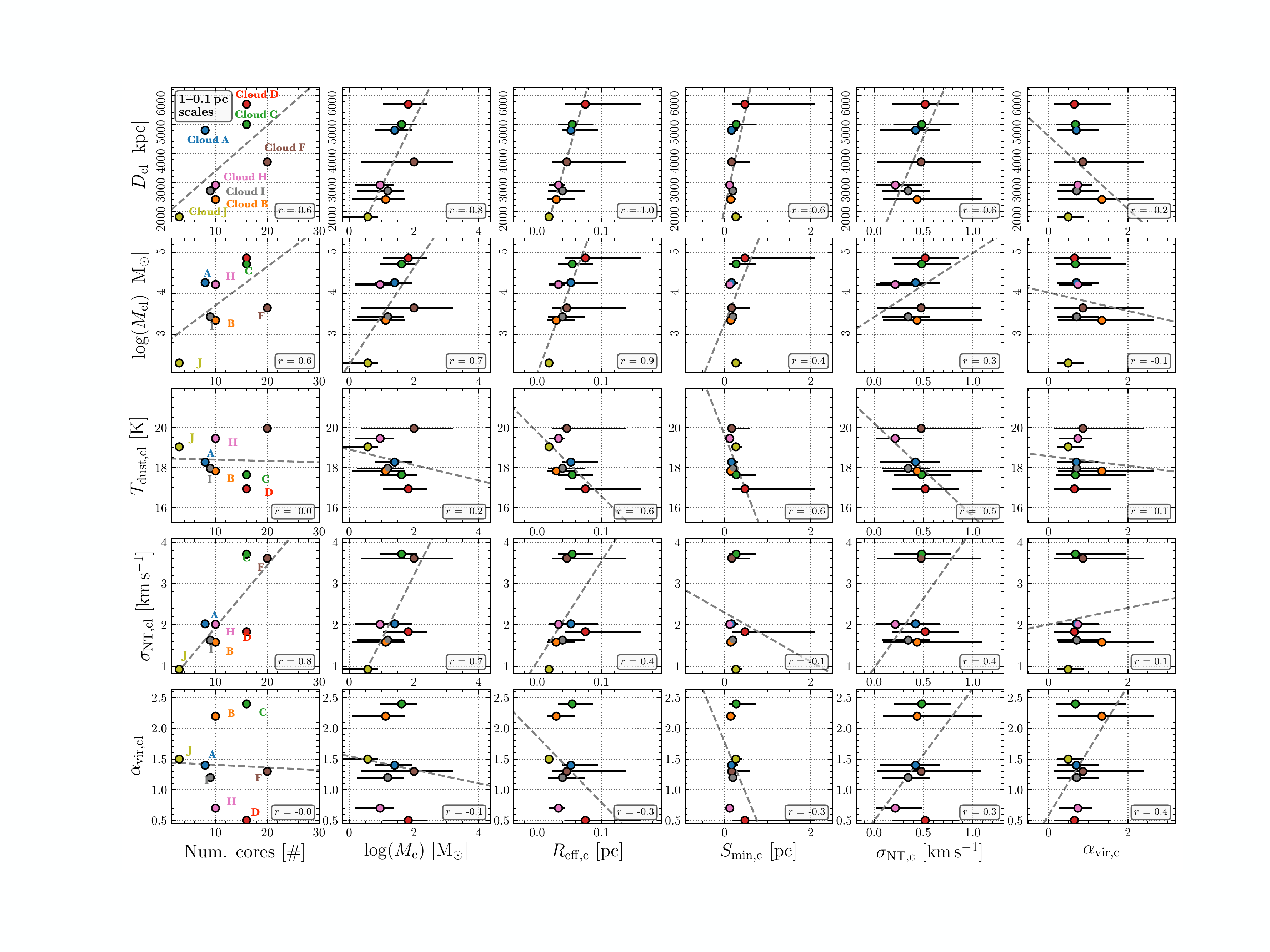}
     	\caption{A comparison between the properties of the ($\sim\,0.1$\,pc scale) cores identified in this work, and their host ($\sim\,1$\,pc scale) cloud properties (see Table\,\ref{tab:cloudprops}). In columns from left to right, we show the number, \note{logarithm of the mass} ($M_\mathrm{c}$), effective radius ($R_\mathrm{eff,c}$), minimum separation ($S_\mathrm{min,c}$), non-thermal velocity dispersion ($\sigma_\mathrm{NT,c}$), and virial parameter ($\alpha_\mathrm{vir,c}$) of the cores within each cloud on the x-axis. In rows from top to bottom, we show the cloud distance ($D_\mathrm{cl}$), \note{logarithm of the mass} ($M_\mathrm{cl}$), mean dust temperature ($T_\mathrm{dust,cl}$), non-thermal velocity dispersion ($\sigma_\mathrm{NT,cl}$), and virial parameter ($\alpha_\mathrm{vir,cl}$) on the y-axis. The points shown in all except the first column represent the mean core properties, and the error bars show the range between the minimum and maximum values of the cores within a given cloud. As highlighted in the upper left panel, the colour of each point corresponds to their host cloud. The dashed grey lines show the linear relation ($y=ax+b$) from the least-squares fit of the variables plotted on the x- and y-axis. The Pearson's-$r$ value are given within the upper right of each panel.}
     	\label{fig:corecloudcomp}
 \end{figure*}

 We now use this hierarchical structure to investigate how the global properties inherited from the IRDC ($\sim$\,1\,pc scales) affect the properties of their $\sim$\,0.1\,pc scale cores identified here, and how the properties of these $\sim$\,0.1\,pc scale cores then influence the smallest $\sim$\,0.01\,pc scale core-fragments into which they fragment.

 We first compare the cloud properties to the properties of the cores identified within this work (Table\,\ref{tab:core_physprops}). The cloud properties are given in Table\,\ref{tab:cloudprops}, which have been taken from \citet[][their Table\,1]{kainulainen_2013}. These authors determine the masses of the clouds from the combined mid- and near-infrared extinction maps (see Figure\,\ref{fig:mom_maps}), and the velocity dispersion from $^{13}$CO emission cubes \citep{jackson_2006}. Also included in Table\,\ref{tab:cloudprops} are the mean {\it Herschel} derived dust temperatures measured within the footprint of the ALMA observations for each cloud, which are used to determine the corresponding non-thermal velocity dispersions and sonic Mach numbers (following section\,\ref{sec:physprops}). 

 Figure\,\ref{fig:corecloudcomp} presents the comparison between the core and cloud properties.\footnote{It is worth noting that Cloud G is not included within this analysis as no continuum cores were identified within the mapped region of this source (section\,\ref{sec:results}). Moreover, following \citet{kainulainen_2013}, Cloud E has been removed from this analysis due to its complex velocity structure observed in the $^{13}$CO observations, and hence the large uncertainty associated with the cloud dynamical properties.} In columns of panels from left to right, we show the number, mass, effective radius, minimum separation, non-thermal velocity dispersion, and virial parameter for the identified cores on the x-axis. In rows of panels from top to bottom, we show how the core properties vary with the cloud distance, mass, dust temperature, non-thermal velocity dispersion, and virial parameter on the y-axis. Where correlations between the core and cloud properties appear to be present, we conduct a least-squares minimisation to a linear relation of $y=ax+b$ to obtain $a$ and $b$ for the plotted variable on the x-axis ($x$) and y-axis ($y$). The results of this fitting routine are shown as a dashed grey line on the panels in Figure\,\ref{fig:corecloudcomp}. We also determine the Pearson's $r$ value for each parameter set where a fit is possible, which is given in the upper right of the panels. Here, an $r$ of 1(-1) indicates a perfect positive(negative) linear relationship between the variables, whilst an $r$ of 0 indicates no linear relationship between variables.

 We first discuss the lack of correlations observed between the cloud and core properties. We find that the non-thermal velocity dispersion and the virial state of the cores do not strongly correlate to any of the compared cloud properties ($|r|<0.6$). This could be a result of using different molecular lines to determine the dynamical properties for the clouds and cores. The clouds were investigated by \citet{kainulainen_2013} using $^{13}$CO(1-0) emission, whereas here for the cores, we make use of N$_{2}$H$^{+}$ emission that has a significantly higher critical density and chemical formation pathway that favours colder, denser gas \citep{caselli_2002,caselli02c,hacar_2011,hacar_2013, henshaw_2014,kauffmann_2017,barnes20a}. Therefore, the cloud and core scale dynamics could originate from very different density and temperature layers within the molecular cloud. Additionally, it is likely that the clouds contain multiple distinct velocity components that are not resolved within the lower density molecular line tracers such as $^{13}$CO(1-0) (e.g. \citealp{henshaw_2013,jimnez-serra_2014,hacar_2016c,barnes_2018}). If not separated, these would artificially increase the measured velocity dispersion and inferred virial parameters of the clouds (e.g. \citealp{henshaw_2014}). Alternatively, this lack of correlation could suggest that the cores are dynamically decoupled from their host clouds (e.g. \citealp{goodman98,hacar_2016a}). In this scenario, we are observing the cores at a stage when a significant fraction of the turbulence initially inherited from the host molecular cloud has been dissipated, hence the cores are unstable to collapse and are doing so faster than the global cloud. The initial physical properties of these cores could be set by the host molecular cloud, yet their dynamics are now independent of their host environment.  
 
 We now discuss the correlations observed between the core and cloud properties. Firstly, we find that the number of cores identified within each cloud, core mass and radius increase with increasing distance and mass of the cloud. We note that these correlation all have a Pearson's $r>0.6$, albeit with a strong dependence on the properties determined for Cloud J. There is no physical reason to expect why these properties should scale with increasing cloud distance. This then could be a resolution and sensitivity effect, whereby more massive cores with larger radii are identified within less resolved clouds found at larger distances. Moreover, at larger distances more projected area of the clouds can be mapped for the same angular area on the sky, and, therefore, there is a higher likelihood of identifying more cores. Alternatively, as the cloud masses are not directly proportional to their distances, and the number of cores, core masses and radii also appear to scale with cloud mass, there is a different simple conclusion that could be drawn from this result: more massive clouds produce more cores, which also are larger and more massive. 

 We see that the minimum separation between the cores correlates to the distance and dust temperature of the clouds ($|r|>0.6$). The correlation to the temperature could be linked to the Jeans fragmentation of the cloud, where higher temperatures produce a larger Jeans length (see equation\,\ref{eq:jeanslength}). Assuming the mean temperature of 18\,K, and inputting the cloud mass and sizes into equation\,\ref{eq:jeanslength} (and a spherical geometry for the density), we calculate $\lambda_\mathrm{J}$ of 0.7\,pc and 3.7\,pc for the lowest and largest mass clouds in the sample, respectively. These values are significantly larger than the distribution of observed minimum separations between the cores. There is then the caveat that the observed core spatial distribution is the two-dimensional projection of the true three-dimensional structure of the cloud, and hence could be strongly dependent on the unknown cloud orientation and internal structure (e.g. \citealp{henshaw_2016c}). 

 Finally, we find that the number of cores identified within each cloud, core mass and radius increase with increasing non-thermal velocity dispersion (or sonic Mach number). Such a trend is expected from turbulent star formation theory, where a greater degree of turbulence produces greater contrasts above the mean density; i.e. seen as cores here (see \citealp{padoan_2014} for review, also see \citealp{Palau2014, fontani_2018}). 

 \begin{figure*}
 \centering
     \includegraphics[width=0.75\textwidth]{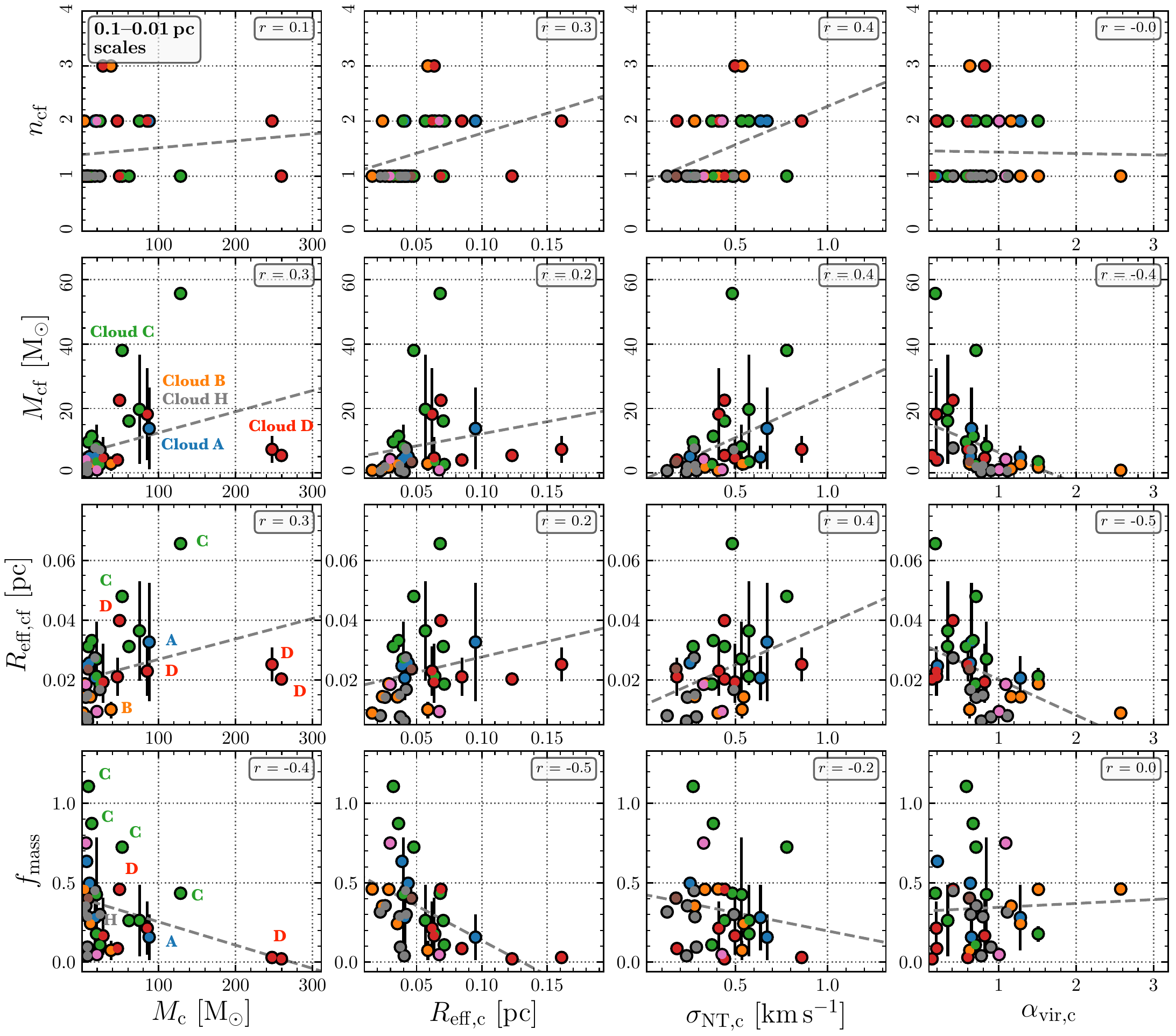}
     	\caption{A comparison between the properties of the ($\sim\,0.1$\,pc scale) cores identified in this work, and the ($\sim\,0.01$\,pc scale) core-fragments they contain (see Table\,\ref{tab:core_physprops} and \ref{tab:core_litprops}). Shown from left to right in columns is the masses ($M_\mathrm{c}$), radii ($R_\mathrm{eff,c}$), non-thermal velocity dispersion ($\sigma_\mathrm{NT,c}$), and virial parameter ($\alpha_\mathrm{vir,c}$) of the cores identified in this work on the x-axis. Shown from top to bottom in rows is the number of core-fragments ($n_\mathrm{cf}$), mass ($M_\mathrm{cf}$), radius ($R_\mathrm{eff,cf}$), and mass fraction ($f_\mathrm{mass}=M_\mathrm{cf}/M_\mathrm{c}$) contained within the core fragment on the y-axis. The points and error bars shown here are the mean and the minimum to maximum value of the fragments contained within each core, respectively. The dashed grey lines shows the linear relation ($y=ax+b$) from the least squares fit of the variables plotted on the x- and y-axis. The Pearson's-$r$ value are given within the upper right of each panel.}
     	\label{fig:corecorecomp}
 \end{figure*}

 \subsubsection{From cores ($\sim$\,0.1\,pc scales) to core-fragments ($\sim$\,0.01\,pc scales)}

 We now compare the properties of the cores identified in this work to the properties of their contained smaller scale core-fragments.\footnote{Cloud F and G are not included in this analysis due to the lack of cross over with the \citetalias{liu_2018} core catalogue, and the catalogue of cores identified in this paper.} Here, we also calculate the fraction of mass contained within each core fragment to the total mass of the host core; i.e. $f_\mathrm{mass}=M_\mathrm{cf}/M_\mathrm{c}$, where $M_\mathrm{c}$ is the host core and $M_\mathrm{cf}$ is the mass of the contained fragment.

 Shown in columns from left to right in Figure\,\ref{fig:corecorecomp} are the number, masses, radii, non-thermal velocity dispersion, and virial parameter of the cores identified in this work on the x-axis. Shown in rows from top to bottom is how these core properties vary with the mass, radius, and mass fraction contained within the core fragment on the y-axis. The points and error bars shown here are the mean and the minimum to maximum value of the fragments contained within each core, respectively. Again here we conduct a least-squares minimisation to a linear relation and determine the Pearson's $r$, which is shown in the upper right of the panels in Figure\,\ref{fig:corecorecomp}. 
 
 We find no significant trends ($|r|>0.6$) between the core properties and number, masses or sizes of their contained core-fragments. Moreover, we find that none of the core properties has any influence on the fraction of the total core mass contained within each core fragment. \note{Interestingly, we then find that the correlations observed on the cloud to core scales are not present on the core to core-fragment scales. \citet{Palau2014} conducted an in-depth study of fragmentation from 0.1 to $<$0.01\,pc scales within a sample of 19 high-mass clumps (also see \citealp{Palau2013, palau15}, and found tentative trends between host core (clump) and their smaller scale fragment(s) properties; e.g. more fragmentation with increased density. Similarly, \citet{fontani_2018}, overall, found weak trends between host clump and core properties from 0.1 to 0.01\,pc scales across a sample of 11 high-mass star-forming regions; e.g. more fragmentation with increased turbulence. The reason for the breakdown in the cloud to core correlations on the core to core-fragment scales here is then not clear. It could be a result of the different methods and tracers used to determine the core and fragment properties. For example, the comparison of datasets including different spatial filtering e.g. due to the use of single-dish observations, or differing molecular line to trace dynamics. Alternatively, the differences in trends could be a result of time variability, where a single time snapshot of a population of cores at different evolutionary stages would then naturally produce scatter.}
 

\section{Conclusions}\label{sec:conclusions}

We have presented 3\,mm wavelength ALMA observations towards 10 infrared dark molecular clouds (IRDCs). This set of observations currently represents the highest resolution ($\sim$\,3\arcsec; $\sim$\,0.05\,pc), highest sensitivity ($\sim$\,0.15\,mJy\,beam$^{-1}$ full bandwidth or $\sim$\,0.2\,K per 0.1\,\kms\ channel) large mosaics (covering parsecs) for a sample of massive molecular clouds. In this work, we conduct an in-depth analysis of the hierarchical structure present within these molecular clouds, and assess the high-mass star-forming potential across fragmentation scales from clouds ($\sim$\,1\,pc), to clumps ($\sim$\,0.5\,pc), to cores ($\sim$\,0.1\,pc) and finally to core-fragments ($\sim$\,0.01\,pc). The main conclusions of this work are summarised below. 

\begin{itemize}
    \item[(i)] We identify 96 cores across the 10 clouds within the ALMA continuum maps, which we calculate have masses of $M=3.4 - 50.9$\sol\ and number densities of $n_\mathrm{H_2} = 4 - 12\times10^5$\,cm$^{-3}$ (ranges are a standard deviation around the median; section\,\ref{sec:physprops}). We determine their dynamical properties from the brightest velocity component observed within the ALMA N$_2$H$^{+}$\,(1-0) emission cubes towards the position of each core. We find sonic (non-thermal) Mach numbers of $\mathcal M_\mathrm{s} = 0.9 - 2.7$, and virial parameters of $\alpha_\mathrm{vir} = 0.3 - 1.3$ (section\,\ref{sec:dynprops}). These results highlight that the cores identified here are dense, gravitationally bound, and dominated by trans-sonic turbulence. 
    
    \item[(ii)] In addition to the cores identified from the ALMA observations presented here, we include a large sample of cores and clumps from the literature that also covers our 10 cloud sample. The properties of these clump/cores are recalculated using the same assumptions of dust opacity, temperature, and gas-to-dust ratio and definition of radius to produce a homogenised catalogue of core properties. We use the fact that this catalogue has been created using observations that cover the same regions at various spatial resolutions to follow the hierarchical structure within each cloud. To do so, we label which cores are co-spatial, and hence form part of the same fragmentary structure (section\,\ref{sec:frag}). We compare this structure for each cloud to mass and density thresholds for massive star formation. We find that from the cloud ($\sim$10\,pc) to clump ($\sim$0.5\,pc), and to the core ($<$0.1\,pc) scales the fragmentation does not follow a simple power-law relation in the mass-size parameter space, which causes different scales within the same cloud to be classified as high- or low-mass star-forming. Caution must then be taken when using density threshold scaling relations to draw conclusions of the high-mass star-forming potential of a core, clump or cloud across any spatial scale.
    
    \item[(iii)]  When assessing the simple mass-size relations, we find that on size scales of <\,0.02\,pc ($\sim$2000\,au) none of the core-fragments appear to contain enough mass to form a high-mass star without additional accretion. However, here we can use the hierarchical structure to determine if any of the larger cores retain enough mass to form a high-mass without further fragmentation. We find that at a size scale of $\sim$0.1\,pc, there is a sample of 19 cores that have masses of >16\,\sol\ without further fragmentation. Out of these, we find that 6 show no signs of active star formation, whilst 13 have signs of active star formation. These high-mass cores contain trans-sonic non-thermal motions (median $\mathcal M_\mathrm{s}$ of 2.4), are predominately kinematically sub-virial (median $\alpha_\mathrm{vir}$ of 0.5), and require moderate magnetic field strengths for support against collapse (median $B$ of 930\,$\mu$\,G). We find that the sizes of these cores are broadly comparable to the predictions from the core accretion theory, based on their host cloud properties \citep{mckee_2003}. However, to ultimately test the different theories of high-mass star formation, further investigation is needed to assess if (any) of these fragment further and/or have signs of multiple sites of lower-mass star formation (e.g. outflows).
    
    \item[(iv)] We investigate what physical and dynamical properties of the cloud ($>$1\,pc scale) are inherited by or influence their smallest scale core-fragments ($\sim$0.01\,pc scale) populations. We find that more massive, and more turbulent clouds make more $\sim$0.1\,pc scale cores. These $\sim$0.1\,pc scale cores also tend to be more massive within the higher-mass, turbulent clouds. We find tentative evidence that these cores then to fragment into more massive $\sim$0.01\,pc scale core-fragments.
\end{itemize}



\section*{Acknowledgements}

\note{We would like to thank the referee for their constructive feedback that helped improve the paper.} ATB and FB would like to acknowledge funding from the European Research Council (ERC) under the European Union’s Horizon 2020 research and innovation programme (grant agreement No.726384/Empire). JEP and PC acknowledge the financial support of the Max Planck Society. IJ-S has received partial support from the Spanish FEDER (project number ESP2017-86582-C4-1-R) and the State Research Agency (AEI; project number PID2019-105552RB-C41). RJP acknowledges support from the Royal Society in the form of a Dorothy Hodgkin Fellowship. ASM research is conducted within the Collaborative Research Centre 956 (sub-project A6), funded by the Deutsche Forschungsgemeinschaft (DFG, project ID 184018867). SF acknowledges the support  the EACOA fellowship from the East Asia Core Observatories Association (EACOA). KW acknowledges support by the National Key Research and Development Program of China (2017YFA0402702, 2019YFA0405100), the National Science Foundation of China (11973013, 11721303), and the starting grant at the Kavli Institute for Astronomy and Astrophysics, Peking University (7101502287). This paper makes use of the following ALMA data: ADS/JAO.ALMA\#2017.1.00687.S \& ADS/JAO.ALMA\#2018.1.00850.S. We would like to thank Audra Hernandez, Vlas Sokolov, and Andy Pon for their input on the proposal for these AMLA observations. ALMA is a partnership of ESO (representing its member states), NSF (USA) and NINS (Japan), together with NRC (Canada), MOST and ASIAA (Taiwan), and KASI (Republic of Korea), in cooperation with the Republic of Chile. The Joint ALMA Observatory is operated by ESO, AUI/NRAO and NAOJ.

\section*{Data availability}

A full machine-readable version of Tables\,\ref{tab:core_obsprops}, \ref{tab:core_physprops} and \ref{tab:core_litprops} is available in the online supplementary of this work. All ALMA observations used within this work are publicly accessible in the ALMA science archive. Additional data products not provided in these tables or available online will be shared on reasonable request to the corresponding author.

\bibliographystyle{mnras}
\bibliography{references}

\appendix

\section{Literature cores and fragmentation analysis}\label{apendix:tables}

\begin{figure*}
\centering
    \includegraphics[width=\textwidth]{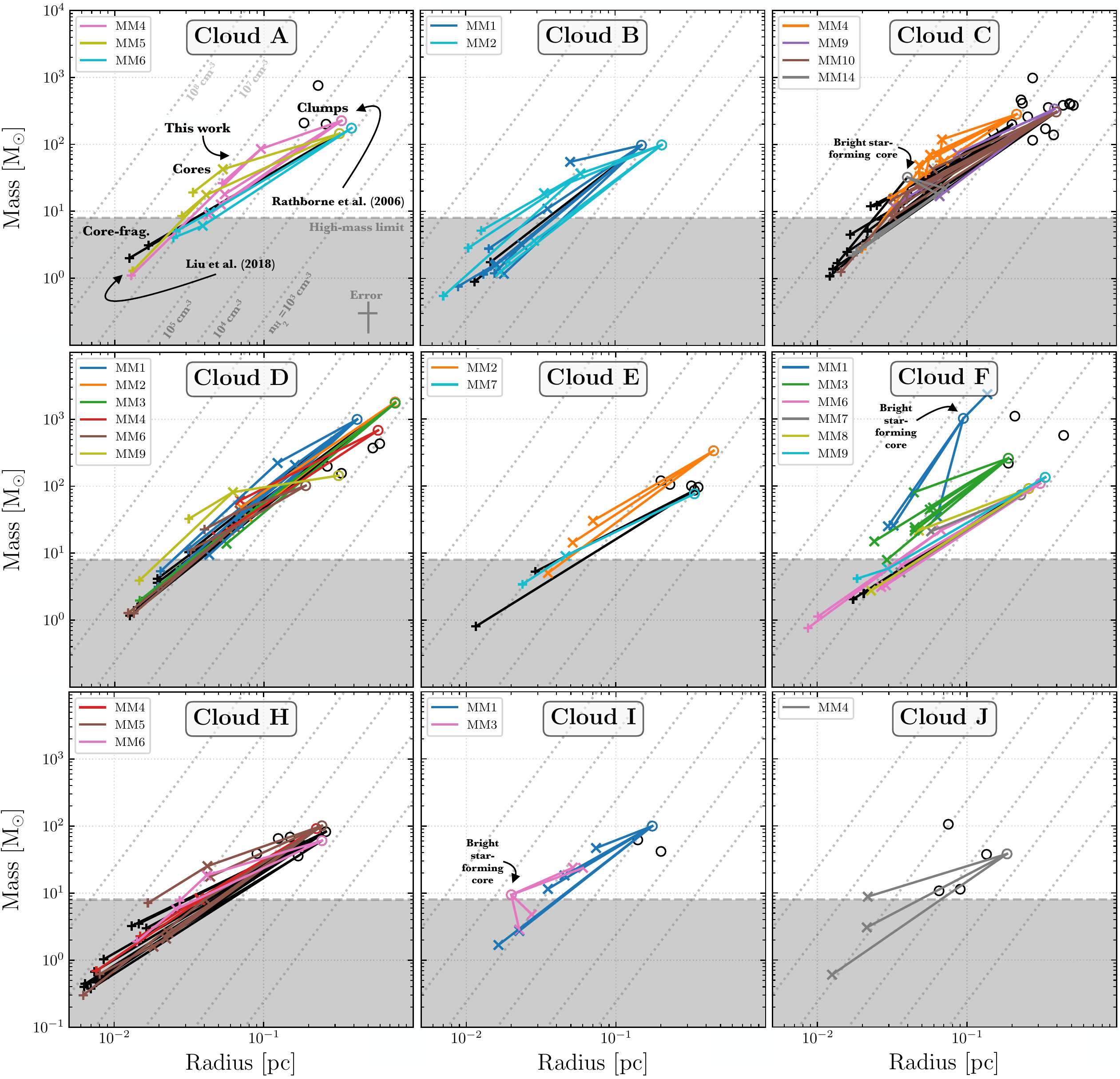}
    	\caption{Mass fragmentation of the cores within each IRDC as a function of the size-scale. The circles, crosses and plus sign markers represent the cores from the homogenised sample identified by \citet{rathborne_2006}, this work and \citet{liu_2018} (section\,\ref{sec:litomogenised}). The straight lines connect the symbols for each core and the larger, host core of which it is a  part (section\,\ref{sec:frag}. The lines and symbols have been coloured by host ``MM'' \citet{rathborne_2006} core, as indicated in the legend located in the upper left (see Table\,\ref{tab:core_litprops}). Also highlighted with labels are the clumps associated within bright star-forming regions, which limits the mass determination from the lower resolution observations from \citet{rathborne_2006}. The resultant large uncertainty on the mass estimate can cause these \citet{rathborne_2006} structures to appear to have masses smaller than is determined in this work. Their hierarchical structure should, therefore, be taken with caution. Cloud G is not shown due to the lack of continuum cores determined in the 3\,mm ALMA continuum observations presented as part of this work (section\,\ref{sec:coreident}). The shaded region shows the 8\,\sol\ mass threshold for a high-mass star. \note{In the lower right corner of the upper left panel, we show a representative uncertainty range $\sim$\,15 and $\sim$\,50\,per cent on the radius and mass, respectively.}} 
    	\label{fig:fragsep}
\end{figure*}

In the section, we present the complete core and homogenised literature core catalogues used throughout this work, which can be found in full, machine-readable format online. Moreover, we present the fragmentation structure determined within each cloud, which has been summarised in Figure\,\ref{fig:fragsep} of the main text (section\,\ref{sec:frag}). 

Table\,\ref{tab:core_obsprops} presents the observed properties of each core determined from the the dendrogram analysis (section\,\ref{sec:coreident}), their spectroscopic properties determined from the N$_2$H$^+$ Gaussian fits (section\,\ref{sec:dynprops}), and host millimetre (MM) core \citep{rathborne_2006}. Table\,\ref{tab:core_physprops} presents the physical properties of the identified cores (sections\,\ref{sec:physprops}, \ref{sec:dynprops}). Table\,\ref{tab:core_litprops} presents the literature core catalogue \citep{rathborne_2006, henshaw_2016c, henshaw_2016d, liu_2018}. Here the spatial (effective) radii and masses of each core have been re-calculated from the observed angular (effective) radii and continuum fluxes using the same set of assumptions for the dust opacity, dust-to-gas ratio, and dust temperature (section\,\ref{sec:litomogenised}). 

\begin{table*}
    \caption{Observational properties of the core population (sections\,\ref{sec:coreident}, \ref{sec:dynprops} and \ref{sec:frag}). Shown in columns are the results from the dendrogram analysis of the core ID and name, the host cloud, the centre RA and Declination, the effective radius ($R_\mathrm{eff}$ in units of arcsec), the total continuum flux density ($S_\nu$) and background subtracted flux density ($S^\mathrm{b}_\nu$), and the peak continuum intensity ($I_\nu$(max)). Also given are the results of the N$_2$H$^+$ Gaussian fits of peak brightness temperature ($T_\mathrm{max}$), centroid velocity ($\mathrm{v}_0$) and velocity dispersion ($\sigma_\mathrm{v}$). We show the millimetre (MM) core in which each core is contained \citep{rathborne_2006}. Finally, we show if the core contains an embedded ({\it Spitzer} or {\it Herschel} 70\,\micron) infrared point source, \note{and the flux density of any associated 70\,\micron\ point sources ($S_\mathrm{70\mu m}$; \citealp{molinari_2016,marton17}).}}
    \label{tab:core_obsprops}
\begin{tabular}{ccccccccccccccc}
\hline \hline
ID & Name & Cloud & RA & Dec. & $R_\mathrm{eff}$ & $S_\nu$ & $S^\mathrm{b}_\nu$ & $I_\nu$(max) & $T_\mathrm{max}$ & $\mathrm{v}_0$ & $\sigma_\mathrm{v}$ & MM core & SF & $S_\mathrm{70\mu m}$\\
 & & & (J2000) $\mathrm{{}^{\circ}}$ & (J2000) $\mathrm{{}^{\circ}}$ & $\mathrm{{}^{\prime\prime}}$ & $\mathrm{mJy}$ & $\mathrm{mJy}$ & $\mathrm{mJy}$ & $\mathrm{K}$ & $\mathrm{km\,s^{-1}}$ & $\mathrm{km\,s^{-1}}$ & $\mathrm{Jy}$ \\
 \hline
1 & A1c1/2 & cloudA & 276.564 & -12.694 & 4.1 & 3.97 & 2.28 & 2.00 & 2.1 & 64.7 & 0.7 & MM4 & y & - \\
2 & A1c3 & cloudA & 276.566 & -12.692 & 2.3 & 0.84 & 0.34 & 0.72 & 2.0 & 65.7 & 0.3 & MM4 & n & - \\
3 & A1c4 & cloudA & 276.564 & -12.692 & 1.8 & 0.41 & 0.06 & 0.46 & 1.9 & 65.0 & 0.5 & MM4 & y & - \\
4 & A1c5 & cloudA & 276.569 & -12.690 & 2.2 & 0.59 & 0.17 & 0.48 & 2.4 & 66.3 & 0.4 & MM4 & y & 3.05 \\
5 & A2c2 & cloudA & 276.580 & -12.688 & 1.9 & 0.45 & 0.17 & 0.58 & 2.4 & 66.3 & 0.4 & MM6 & n & - \\
\dots & \dots & \dots & \dots & \dots & \dots & \dots & \dots & \dots & \dots & \dots & \dots & \dots \\
\hline \hline
\end{tabular}
    \begin{minipage}{0.82\textwidth}
    \vspace{1mm}
    {\bf Note}: The full, machine-readable version of this Table can be obtained from the supplementary online material.
    \end{minipage}
\end{table*}

\begin{table*}
    \caption{Physical properties of the core population (sections\,\ref{sec:physprops}, \ref{sec:dynprops}). Shown in columns are the core name (Table\,\ref{tab:core_obsprops}), the effective radius (in units of parsec; $R_\mathrm{eff}$), the minimum separation or nearest neighbour distance ($S_\mathrm{min}$), the mean dust temperature ($T_\mathrm{dust}$), the mass determined using the mean dust temperature ($M$), the background subtracted mass using the mean dust temperature ($M^\mathrm{b}$), the mass determined using a constant temperature of 18\,K ($M_\mathrm{18K}$), the mass determined from the near- and mid- infrared extinction maps ($M_\mathrm{ext}$; \citealp{kainulainen_2013}),  \note{the mass estimates using a temperature determined from 70\micron\ emission ($M_\mathrm{70\mu m}$)}, the mean density ($n_\mathrm{H_2}$), the free-fall time ($t_\mathrm{ff}$), virial parameter ($\alpha_\mathrm{vir}$), background subtracted virial parameter ($\alpha_\mathrm{vir}^\mathrm{b}$), the non-thermal velocity dispersion ($\sigma_\mathrm{NT}$), the sonic Mach number ($\mathcal{M}_\mathrm{s}$), the Alfv\'en Mach number ($\mathcal M_\mathrm{A}$) and magnetic field strength to reach gravitational equilibrium ($B$; section\,\ref{sec:stability_thermturbmag}).}
    \label{tab:core_physprops}
\begin{tabular}{cccccccccccccccccc}
\hline \hline
 Name & $R_\mathrm{eff}$ & $S_\mathrm{min}$ & $T_\mathrm{dust}$ & $M$ & $M^\mathrm{b}$ & $M_\mathrm{18K}$ & $M_\mathrm{ext}$ & $M_\mathrm{70\mu m}$ & $n_\mathrm{H_2}$ & $t_\mathrm{ff}$ & $\alpha_\mathrm{vir}$ & $\alpha_\mathrm{vir}^\mathrm{b}$ & $\sigma_\mathrm{NT}$ & $\mathcal{M}_\mathrm{s}$ & $\mathcal{M}_\mathrm{A}$ & $B$ \\
 & $\mathrm{pc}$ & $\mathrm{pc}$ & $\mathrm{K}$ & $\mathrm{M_{\odot}}$ & $\mathrm{M_{\odot}}$ & $\mathrm{M_{\odot}}$ & $\mathrm{M_{\odot}}$ & $\mathrm{M_{\odot}}$ & $\mathrm{10^5\,cm^{3}}$ & $\mathrm{10^4\,yr}$ &  &  & $\mathrm{km\,s^{-1}}$ & & & $\mathrm{\mu G}$ \\ \hline
A1c1/2 & 0.095 & 0.2 & 17.6 & 87.6 & 50.2 & 85.3 & 45.0 & - & 3.5 & 5.2 & 0.6 & 1.1 & 0.7 & 2.8 & 6.1 & 665 \\
A1c3 & 0.055 & 0.2 & 17.4 & 18.6 & 7.7 & 17.9 & 14.3 & - & 4.0 & 4.9 & 0.2 & 0.5 & 0.1 & 0.3 & 4.3 & 492 \\
A1c4 & 0.043 & 0.2 & 17.5 & 9.1 & 1.4 & 8.8 & 4.2 & - & 4.1 & 4.8 & 1.3 & 8.3 & 0.4 & 1.7 & 2.1 & 251 \\
A1c5 & 0.051 & 0.3 & 18.0 & 12.6 & 3.5 & 12.6 & - & 8.3 & 3.2 & 5.4 & 0.6 & 2.1 & 0.2 & 1.0 & 3.2 & 335 \\
A2c2 & 0.044 & 0.1 & 17.3 & 10.1 & 3.7 & 9.7 & 9.9 & - & 4.2 & 4.7 & 0.6 & 1.7 & 0.3 & 1.1 & 3.1 & 368 \\
\dots & \dots & \dots & \dots & \dots & \dots & \dots & \dots & \dots & \dots & \dots & \dots & \dots & \dots \\
\hline  \hline
\end{tabular}
    \begin{minipage}{0.9\textwidth}
    \vspace{1mm}
    {\bf Note}: The full, machine-readable version of this Table can be obtained from the supplementary online material. Were applicable, the online table also includes all the properties determined with and without background subtraction for the flux density, and using the measured mean and constant 18\,K dust temperature.
    \end{minipage}
\end{table*}

\begin{table*}
    \caption{Properties of the homogenised literature core sample (section\,\ref{sec:litomogenised}). Shown in columns is the host cloud name and ID, the centre RA and Declination, the effective radius in arcsec and parsec ($R_\mathrm{eff}$), the total flux density at the observed frequency ($S_\lambda$), the mass assuming a constant temperature of 18\,K, the wavelength of the observations (see Table\,\ref{tab:lit_obsprops}), and the reference \citep{rathborne_2006, henshaw_2016c, henshaw_2016d, liu_2018}.}
    \label{tab:core_litprops}
\begin{tabular}{ccccccccccc}
\hline \hline
Cloud & ID  & Core & RA & Dec. & $R_\mathrm{eff}$ & $R_\mathrm{eff}$ & $S_\lambda$ & $M_\mathrm{18K}$ & $\lambda$ & Reference \\
 &  &  & $\mathrm{{}^{\circ}}$ & $\mathrm{{}^{\circ}}$ & $\mathrm{{}^{\prime\prime}}$ & $\mathrm{pc}$ & $\mathrm{Jy}$ & $\mathrm{M_{\odot}}$ & $\mathrm{mm}$ &  \\
 \hline
cloudA & G018.82-00.28 & A3c4 & 276.5900 & -12.6863 & 0.72 & 0.017 & 3.79 & 3.09 & 1.3 & \citet{liu_2018} \\
cloudA & G018.82-00.28 & MM5 & 276.5875 & -12.6864 & 15.00 & 0.320 & 234.60 & 144.62 & 1.3 & \citet{rathborne_2006} \\
cloudA & G018.82-00.28 & A1c3 & 276.5662 & -12.6921 & 2.34 & 0.055 & 0.84 & 17.95 & 3.0 & This work \\
cloudA & G018.82-00.28 & A2c1 & 276.5777 & -12.6883 & 1.06 & 0.025 & 4.91 & 4.01 & 1.3 & \citet{liu_2018} \\
cloudA & G018.82-00.28 & A1c2 & 276.5640 & -12.6937 & 2.26 & 0.052 & 32.38 & 26.43 & 1.3 & \citet{liu_2018} \\
cloudA & G018.82-00.28 & A1c5 & 276.5692 & -12.6900 & 2.20 & 0.051 & 0.59 & 12.61 & 3.0 & This work \\
\dots & \dots & \dots & \dots & \dots & \dots & \dots & \dots & \dots & \dots \\
 \hline  \hline
\end{tabular}
    \begin{minipage}{0.84\textwidth}
    \vspace{1mm}
    {\bf Note}: The full, machine-readable version of this Table can be obtained from the supplementary online material. This is a literature compilation, please ensure to cite each individual study when making use of the contents of this table.
    \end{minipage}
\end{table*}

\section{Comparison of feathering and uv-combination}\label{apendix:cleanfeather}

The combination of multiple datasets is typical for interferometric observations, where the range of recoverable spatial scales is limited by the baselines included in each observation. There is still, however, much debate within the literature for the best practises for this combination, and particularly for the case where single-dish observations are used with interferometric datasets to recover the zero-spacing. Throughout this work, we make use of ALMA observations taken with both the 12m and 7m arrays, which were reduced using the {\sc casa-pipeline} (version: 5.4.0-70) and combined using the {\sc feather} (CASA version: 4.7.0). For the follow-up work investigating the nitrogen fractionation, we have also combined the single-dish observations with the 12m and 7m images using the same procedure \citep{fontani_2020inprep}. We chose to feather the images produced directly by the pipeline for convenience. However, an alternative, commonly used approach, is to combine the array configurations in the {\it uv}-plane and to then image them together using the e.g. {\sc clean} function.

In light of the above, in this section, we conduct a comparison between the continuum map and \ntwohoz\ cube within Cloud C produced by combining the 12\,m and 7\,m datasets in the {\sc feather} and {\sc tclean} functions. Firstly, for the continuum, we use the calibrated measurement sets produced by the pipeline that have not been continuum subtracted and use the line-free parts of the bandwidth identified from the {\sc hif$\_$findcont} task in the {\sc casa-pipeline}. When imaging with {\sc tclean} (version: 5.6.0), we use natural weighting, a multi-term (multiscale) multi-frequency synthesis deconvolver ({\sc mtmfs} option), and set a high number of iterations to achieve a noise threshold of 0.5\,mJy per beam within a mask that is automatically determined after each minor cycle (see the {\sc auto-multithresh} option for the {\sc usemask} parameter in {\sc tclean}; \citealp{kepley_2020}). Second, for the \ntwohoz\ cubes, we make use of the continuum subtracted measurement set produced by the pipeline. When imaging, we again use natural weighting, a multiscale deconvolver ({\sc multiscale} option), set a high number of interactions to achieve a noise threshold of 21\,mJy per beam per channel within an automatically determined mask.

\begin{figure*}
\centering
    \includegraphics[width=\textwidth]{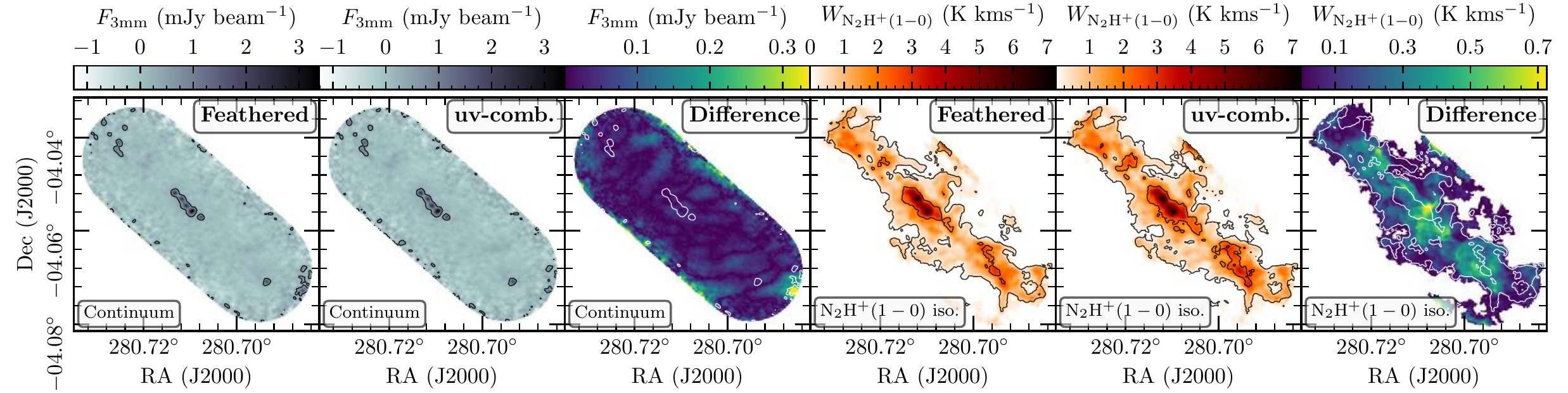}
    	\caption{A comparison between the images produced when feathering the pipeline imaged 12\,m and 7\,m datasets, and when imaging the 12\,m and 7\,m datasets together in the {\it uv}-plane with the clean function for Cloud C (see Figure\,\ref{fig:mom_maps}). The left two panels show the feathered, and {\it uv}-combined, cleaned continuum maps, overlaid with a solid black contour of the 0.48\,mJy\,beam$^{-1}$. The centre-left shows a map of the difference between these two continuum images. The fourth and fifth panels show the feathered and cleaned \ntwohoz\ integrated intensity maps. In both cases the integrated intensity has been determined over the isolated hyperfine component of \ntwohoz, using the same mask and velocity range. Overlaid are solid black contours at signal-to-noise levels of 3 and 5$\sigma$. The rightmost panel shows the difference between these \ntwohoz\ integrated intensity maps. Note that the feathered and cleaned continuum and integrated intensity maps have the same colour scale, and the respective difference map has a range of 1/10 this colour scale.}
    	\label{fig:cleanmaps}
\end{figure*}

Figure\,\ref{fig:cleanmaps} shows the feathered and uv-combined, cleaned continuum maps and \ntwohoz\ integrated intensity maps for Cloud C, where the integrated intensity has been determined using the same mask and velocity range for both cubes. Here we match the colour bar scales for both maps for ease of comparison, and overlay contours at signal-to-noise levels of 3 and 5$\sigma$ (see table\,\ref{tab:obsprops}). We see here that both qualitatively and quantitatively the maps produced with the {\sc feather} and {\sc tclean} functions are very similar, and only minor differences can be seen on close inspection. To further quantify this, in Figure\,\ref{fig:cleanmaps} we also present maps of the absolute difference between the feathered and cleaned images. We find absolute differences of up to $10$\,per cent between the two methods; note the colour scale range used to show the difference maps is 1/10 of the maximum of the cleaned and feathered maps. This difference is, however, small compared to the underlying systematic uncertainties inherent in the physical properties that these maps are used to calculate within this work (e.g. for the mass, where the combined uncertainty from the temperature and distance will be factors of a few higher).

\begin{figure}
\centering
    \includegraphics[width=\columnwidth]{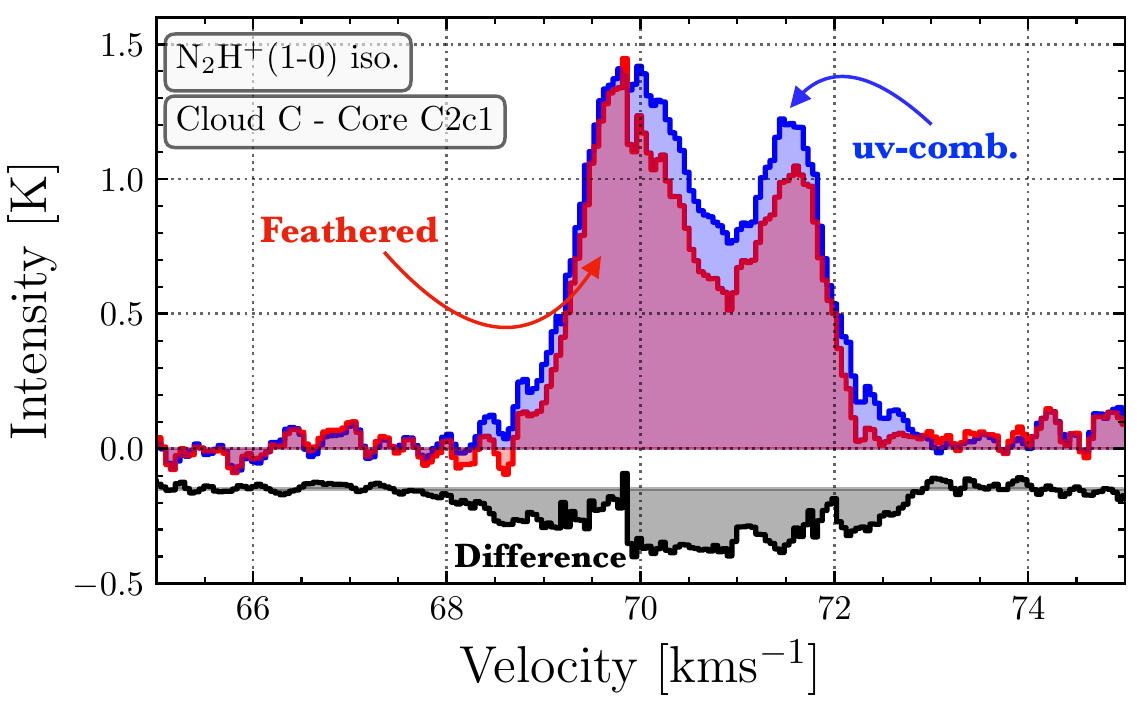}
    	\caption{A comparison between the \ntwohoz\ cubes produced when feathering the pipeline imaged 12\,m and 7\,m datasets, and when imaging the 12\,m and 7\,m datasets together in the {\it uv}-plane with the clean function for Cloud C (see Figure\,\ref{fig:mom_maps}). Shown in blue and red are mean spectra of the isolated hyperfine component of \ntwohoz\ across the core C2c1. Shown in black and centred on -0.15\,K is the difference between these two spectra.} 
    	\label{fig:cleanspec}
\end{figure}

Along with examining the two-dimensional distributions for \ntwohoz, we can also make use of the cubes to compare the {\sc feather} and {\sc tclean} 12m and 7m combination for each velocity slice. Figure\,\ref{fig:cleanspec} shows the spectra of the isolated component of \ntwohoz\ averaged over the core C2c1, which is the closest core to the largest difference in the \ntwohoz\ integrated intensity maps (Figure\,\ref{fig:cleanmaps}). These spectra both show profiles that contain only two Gaussian profiles, which are separated by $\sim$2\,\kms. However, we see that the feathered spectrum has systematically lower intensities than the cleaned spectrum, which is highlighted by the difference profile. Where present, we find that this difference ranges from $20-50$\,per cent of the feathered spectrum intensity, and is, therefore, larger than observed within the integrated intensity map. That said, it is worth keeping in mind that in this work we only use the line-width for the dynamical analysis of the cores, which will be less sensitive to systematic differences within the intensity. 

In summary, in this section, we find that there are some differences between the map produced when feathering the pipeline imaged 12\,m and 7\,m datasets, and when imaging the 12\,m and 7\,m datasets together in the clean function. These differences are of the order $10$\,per cent, but can be more substantial when inspecting the individual slices of a datacube. However, ultimately, the differences between the two methods are small compared to the systematic uncertainty inherent within the physical properties calculated within this work. The analysis presented here then validates our choice of using the feathered, pipeline reduced images, throughout this work.

\label{lastpage}

\end{document}